\documentclass[11pt, oneside]{article}   	
\usepackage{geometry}                		
\usepackage{graphicx}				

\usepackage{latexsym,amsmath,amssymb, amscd,amsfonts, amsthm,epsfig,color,times,appendix,bm}
\usepackage{graphicx}
\usepackage{pictex}
\usepackage{bbold}
\usepackage[english]{babel}
\usepackage{cite}
\newtheorem{theorem}{\bf Theorem}[]

\newtheorem{prop}[]{\bf Proposition}

\newtheorem{lem}[]{\bf Lemma}

\DeclareMathOperator*{\argmin}{arg\,min}

\begin{document}

\title{Minimax estimation of qubit states with Bures risk}

\author{Anirudh Acharya and M\u{a}d\u{a}lin Gu\c{t}\u{a}\\[4mm]
School of Mathematical Sciences, University of Nottingham,\\ University Park, NG7 2RD Nottingham, UK}

\date{}
\maketitle

\begin{abstract}


\noindent 
The central problem of quantum statistics is to devise measurement schemes for the estimation of an unknown state, given an ensemble of  $n$ independent identically prepared systems. For locally quadratic loss functions, the risk of standard procedures has the usual scaling of $1/n$. However, it has been noticed that for fidelity based metrics such as the Bures distance, the risk of conventional (non-adaptive) qubit tomography schemes scales as $1/\sqrt{n}$ for states close to the boundary of the Bloch sphere. Several proposed estimators  appear to improve this scaling, and our goal is to analyse the problem from the perspective of the maximum risk over all states. 

\vspace{2mm}

\noindent
We propose qubit estimation strategies based on separate and adaptive measurements, that achieve $1/n$ scalings for the maximum Bures risk. The estimator involving local measurements uses a fixed fraction of the available resource $n$ to estimate the Bloch vector direction; the length of the Bloch vector is then estimated from the remaining copies by measuring in the estimator eigenbasis. The estimator based on collective measurements uses local asymptotic normality techniques which allows us derive upper and lower bounds to its maximum Bures risk. We also discuss how to construct a minimax optimal estimator in this setup. Finally, we consider quantum relative entropy and show that the risk of the estimator based on collective measurements achieves a rate $O(n^{-1}\log n)$ under this loss function. Furthermore, we show that no estimator can achieve faster rates, in particular the `standard' rate $n^{-1}$.

\end{abstract}

\section{Introduction}

Quantum state tomography plays in an important role in quantum information processing tasks, as a method of estimating and validating experimentally prepared quantum states \cite{MIT}. The aim of state tomography is the estimation of an unknown density matrix $\rho$ from the outcomes of measurements performed on $n$ identical copies of the state available as a resource. The quality of the resulting estimate $\hat{\rho}$ is quantified in terms of its average error, or risk. Given a measurement design $M$, and the corresponding set of outcomes $X$, the risk of the measurement-estimator pair is   
\begin{equation}
R(\rho, \hat{\rho}) := \mathbb{E}\left[ D(\hat{\rho}(X), \rho) \right],
\end{equation}

where the expectation is taken with respect to the measurement outcomes $X$, given the unknown state $\rho$. The risk depends on the choice of the error, or loss function $D(\hat{\rho},\rho)$, which is a measure of the deviation of the estimated state from the true state $\rho$. Examples of commonly used loss functions are squared Frobenius (norm-two) distance,  infidelity, the trace-norm distance, and the Bures distance. The risk is a function of the resource size $n$, and one is interested in its behaviour in the limit of large $n$. Typically, for a `good' estimator and particular choices of loss functions, (e.g. locally quadratic functions) the optimal risk exhibits a rate of $O(1/n)$ \emph{uniformly} over all states $\rho$.   

However, for certain loss functions (e.g. squared Bures distance, or infidelity) , the risk is known to behave differently for states of unknown purity \cite{MahlerRozema,AcharyaGuta}. This is readily illustrated in the qubit case, where the fidelity between the state $\rho$ and its estimate $\hat{\rho}$ is defined as $F(\rho, \hat{\rho}):= [{\rm Tr}(\sqrt{\sqrt{\rho} \hat\rho \sqrt{\rho}})]^2$ and can be expressed as 
\begin{equation*}
F(\rho, \hat{\rho}) := \frac{1}{2}(1+ \sqrt{1-\vert r \vert} \sqrt{1-\vert \hat{r} \vert} +\boldsymbol{r}\cdot\hat{\boldsymbol{r}}),
\end{equation*}
where $\boldsymbol{r}, \hat{\boldsymbol{r}} \in \mathbb{R}^3$ are the  Bloch vectors of the two states. For $\rho$ within the Bloch sphere, the fidelity is locally quadratic in the components  $\delta_i :=  r_i - \hat{r}_i $, with $i=x,y,z$. However, for states close to the boundary of the Bloch sphere, the fidelity becomes linear in $| \delta_i |$. Standard tomographic estimation of the Bloch vector components by measuring the spin operators $\sigma_x, \sigma_y, \sigma_z$ gives an accuracy of the order of $n^{-1/2}$ in estimating $\delta_i$. This implies that for a loss function such as the infidelity $1-F(\hat{\rho}, \rho)$, or the square Bures distance 
$D_B(\rho, \hat{\rho}_n)^2 := 2(1 - \sqrt{F(\rho, \hat{\rho}_n)})$, the risk scales as $O(1/n)$ for states within the Bloch sphere, but only as $O(1/\sqrt{n})$ for nearly pure states.

This poor scaling  for nearly pure states is significant as the preparation and estimation of pure states is ubiquitous in quantum information processing tasks. Although many papers discuss the issues surrounding quantum tomography for Bures risk, we consider the problem in the context of minimax estimation (see also \cite{FerrieBK2,FerrieBK}), i.e. where the figure of merit is the \emph{maximum risk} over all states  
\begin{equation}\label{eq:maximumrisk}
R_{max} (\hat{\rho}_n):= \sup_{\rho}  R(\rho, \hat{\rho}_n) = \sup_{\rho}\mathbb{E}\left[ D_B(\rho, \hat{\rho}_n)^2 \right] 
\end{equation} 
Our aim is to show that \emph{adaptive, separate} measurement strategies can achieve the $n^{-1}$ scaling of the maximum risk. We also consider collective measurements, and derive an upper bound to the asymptotic constant of the maximum risk. Our analysis shows that the problem of finding \emph{minimax} estimators, i.e. estimators with smallest possible asymptotic constant, reduces to that of finding minimax estimator for the `classical' problem of estimating the coin toss probability with respect to the square Hellinger distance risk.

Several estimation methods have been proposed in the literature, involving both global and local measurement strategies, with the aim of improving the poor scaling of fidelity based risks for nearly pure states. However, to the best of our knowledge a scaling of $n^{-1}$ of the maximum risk has not been demonstrated for any of these estimators.  Two-steps adaptive quantum tomography \cite{GillMassar,RehacekEnglert,BaganBallesterGill,MahlerRozema,HouZhu} involves using a fraction $n_1$ of the available resource $n$ to obtain a preliminary estimate of the eigenbasis of $\rho$, and then performing measurements along the estimated eigenbasis on the remaining $n-n_1$ copies of the state. In \cite{BaganBallesterGill} it is shown that using a vanishing fraction $n_1 = n^{\alpha}$, with $1/2 < \alpha < 1$ for the preliminary estimate gives a rate of $O(1/n)$ for the average infidelity with respect to certain distributions over states. However, it has been pointed out in \cite{MahlerRozema}, that for certain states a vanishing fraction is insufficient, and that for almost pure states the worst case infidelity scales as $O(n^{-5/6})$. Numerical results in \cite{MahlerRozema} suggest that using a fixed fraction $n_0 = \beta n$ instead gives the $O(1/n)$ scaling for nearly pure states. The two step adaptive protocols have been experimentally implemented \cite{MahlerRozema,HouZhu}, showing a quadratic improvement in scaling for nearly pure states. The extension of the two-step adaptive protocol to a fully adaptive one has been considered in \cite{RehacekEnglert,OkamotoLefuji,Fujiwara,Ferrie}, where the measurement basis is aligned according to a current estimate after every measurement step. In the Bayesian framework, `self-learning' measurement protocols have been considered in \cite{FischerKienle,GranadeFerrie,HannemannReiss,KravtsovStraupe}. A detailed review of various adaptive protocols and experimental results is found in \cite{Straupe}.

Protocols considering collective (or joint rather than separate) measurements  have also been considered \cite{BaganBallester,BaganBaig,GutaJanssens,GutaKahn}. It is known that joint measurements perform better than separate measurements in the case of mixed states \cite{BaganBaig}. In a Bayesian framework, \cite{BaganBallester} showed that with certain optimal joint measurements, the asymptotic infidelity averaged over a prior distribution achieved a value of $\frac{3+2\langle r \rangle}{4n}$ for mixed qubit states, where $\langle r \rangle$ is the mean purity over the prior distribution. 
Work in \cite{GutaJanssens} proposes a two-step adaptive estimation strategy that is shown to be \emph{locally} optimal, achieving an infidelity risk of $\frac{1+4\lambda_{\text{max}}(\rho)}{4n}$ for mixed qubit states. However these theoretical results cannot be directly used to derive the $n^{-1}$ scaling of the minimax risk.



\vspace{2mm}


We propose two different estimators, one based on adaptive local measurements similar to \cite{BaganBallesterGill, MahlerRozema}, and a second based on global collective measurements and LAN as in \cite{GutaJanssens,GutaKahn}.

In terms of local measurements, we consider a two step adaptive strategy much in line with already proposed estimators \cite{GillMassar,RehacekEnglert,BaganBallesterGill,MahlerRozema,HouZhu}. A \textit{fixed fraction} of the total sample size $n$ is used to obtain a preliminary estimate $\mathbf{\tilde{r}}$ of the Bloch vector $\mathbf{r}$ of the state, by performing standard tomographic measurements of the spin observables. The remaining copies of the state are used to estimate the eigenvalues of the state by performing measurements along the estimated direction $\mathbf{\tilde{r}}/\vert \mathbf{\tilde{r}} \vert$. The final estimate $\hat{\rho}_n$ of the state is then constructed from the estimated eigenvalues in the adaptive step and the preliminary estimate. For this estimator, we upper-bound the maximum Bures risk and demonstrate a scaling of $n^{-1}$.

The estimator based on global collective measurements uses established LAN results for qubit states \cite{GutaJanssens,GutaKahn}. The measurement strategy involves two stages. The first stage involves the standard tomographic measurements of the spin components on a \textit{vanishing number} of copies of the state $\tilde{n} \ll n$. A preliminary estimate $\tilde{\rho}$ is constructed from the outcomes. 
The second measurement stage depends on this preliminary estimate, and for technical purposes related to the asymptotic analysis we consider the following two cases: $|\tilde{\bf r}|<\delta$ and  $|\tilde{\bf r}|\geq \delta$ for some small constant $\delta>0$. 

When $\tilde{\rho}$ is close to the fully mixed state, the standard tomographic measurements are performed on the remaining copies of the state. 
When the preliminary estimate $\tilde{\rho}$ is away from the fully mixed state, a joint measurement is performed on the remaining copies of the state. The joint state $\rho^{\boldsymbol{\theta}}_n := \rho^{\otimes n}_{\boldsymbol{\theta}}$ has a block-diagonal form following the Weyl decomposition of the underlying space $(\mathbb{C}^2)^{\otimes n}$. Information about the eigenvalue parameter $\lambda$ is encoded in a probability distribution over the different blocks of the decomposition, while information about the local parameters $(u,v)$ is encoded in the block states. We consider a parameterisation of states $\rho_{\boldsymbol{\theta}}$, with $\boldsymbol{\theta} = (\lambda, u,v) \in \mathbb{R}^3$, where $\lambda$ parametrises the smallest eigenvalue of the states and $\boldsymbol{w}=(u,v)$ are certain local rotation parameters around a fixed state $\rho_0$. The parameter $\lambda$ is then estimated from the outcomes of a ``which-block'' measurement, while the local parameters $u, v$ are optimally estimated by exploiting the local asymptotic normality of the block states. The LAN results in \cite{GutaJanssens} establish that in the limit $n \rightarrow \infty$, the block states converge to Gaussian states $\phi^{\boldsymbol{w}}$, with displacement proportional to parameters $(u,v)$ (Theorem \ref{le:LAN}). The optimal estimator of $u,v$ is then the optimal estimator of the displacement of a Gaussian state $\phi^{\boldsymbol{w}}$, which is known to be the heterodyne measurement. 
We derive minimax upper and lower bounds for the risk (\ref{eq:maximumrisk}), and demonstrate that the maximum Bures risk for the estimator $\hat{\rho}_n$ scales as $C/n$, with $C$ being a constant. We obtain lower bound of $5/4$ for this constant, and an upper bound of $3/2$.

An important element in the derivation of the upper-bounds for the maximum risk of both estimators is the fact that Bures distance can locally approximated as a sum of contributions from the eigenvalue parameter and the `rotation' parameters. More explicitly, we have that
 
\begin{equation*}
 D_B(\rho,\hat{\rho}_n)^2  \approx D_H(\boldsymbol{\lambda},\hat{\boldsymbol{\lambda}})^2  + \frac{1}{4} \frac{(1-2\lambda)(1-2\hat{\lambda})}{\sqrt{(1-\lambda)(1-\hat{\lambda})} 
+ \sqrt{\lambda \hat{\lambda}}}  \Phi^2 ,
\end{equation*}

where $D_H(\boldsymbol{\lambda}, \hat{\boldsymbol{\lambda}})^2 := \Vert \sqrt{\boldsymbol{\lambda}} - \sqrt{\hat{\boldsymbol{\lambda}}} \Vert^2 $ with $\boldsymbol{\lambda} = (\lambda , 1- \lambda)^{T}$ is the classical Hellinger distance, and $\Phi$ is the angle between the Bloch vectors of the two states. The optimal rate of estimation of the `rotation' parameters is easily shown to be $1/n$ for all states.  The problem of establishing minimax results for the Bures distance therefore converts a problem of establishing minimax results for the Hellinger risk of estimating the eigenvalue parameter. To the best of our knowledge such minimax results for the Hellinger risk are not known. Instead, we upper-bound this risk by the Kullback-Leibler risk, and use known results about the minimax estimator in this case. However, in section \ref{sec:minimax} we propose that a minimax optimal estimator of the classical parameter $\lambda$ under the Hellinger risk gives a minimax optimal estimator $\hat{\rho}_n$ for qubit states.

In section \ref{sec:QRE}, we consider the quantum relative entropy (QRE) $S(\rho \Vert \rho^{\prime}) = {\rm Tr}[ \rho( \log{\rho} - \log{\rho^{\prime}})]$, and bound the maximum risk under this loss function. As in the case with the Bures distance, an important element is a local decomposition of the QRE risk into contributions from the Kullback-Leibler risk and a term involving the `rotation parameters'. Namely, we have that for qubit states 
\begin{equation*}
S(\rho \Vert \hat{\rho}) = D_{KL}(\boldsymbol{\lambda}, \hat{\boldsymbol{\lambda}}) +  
 \frac{1-2\lambda }{4} (\Phi^2 + O(\Phi^4) ) \log{\left( \frac{1-\hat{\lambda}}{\hat{\lambda}} 
\right)}, 
\end{equation*}

where $D_{KL}(\boldsymbol{\lambda}, \hat{\boldsymbol{\lambda}})$ is the Kullback-Leibler distance between the two distributions $\boldsymbol{\lambda} = (\lambda, 1-\lambda)$ and $\hat{\boldsymbol{\lambda}} =(\hat{\lambda},1-\hat{\lambda})$. Using this local decomposition we show that the global estimator we propose achieves a rate of $O(n^{-1} \log{n})$. Furthermore we make the case that no estimator can achieve faster rates, and show that the minimax QRE risk scales as $O(n^{-1} \log{n})$.

The paper is organised as follows. In section $\ref{sec:localmeasurement}$ we consider an estimator based on local measurements and detail a two step adaptive measurement strategy. We demonstrate that the proposed estimator achieves a minimax rate of $1/n$. In section \ref{sec:globalmeasurement} we propose a second estimator based on global collective measurements. We begin in section \ref{sec:parameterisation} by describing the preliminary measurement stage and introduce our parametrisation of states. In section \ref{sec:jointmeasurement}, we describe the second measurement stage, and overview the block decomposition of the joint state and results of LAN. The minimax bounds for this estimator are derived in section \ref{sec:bounds}, and in section \ref{sec:minimax} we discuss and state the proposition that a minimax estimator for the Hellinger loss function implies a minimax optimal estimator for qubit states. Finally in section \ref{sec:QRE} we consider the quantum relative entropy and establish that the minimax rate under this loss function scales as $O(n^{-1} \log{n})$.

\section{Estimator based on local adaptive measurements}\label{sec:localmeasurement}

We let $\rho$ be an arbitrary density matrix associated with a single qubit state. Given $n$ identical copies of the state as a resource, we wish to construct an estimator of the state. As briefly discussed in the introduction, in this section we propose a two-step adaptive measurement strategy based on local measurements. While the idea of an adaptive local measurement strategy is not new and has been treated in various instances in the literature \cite{GillMassar,RehacekEnglert,BaganBallesterGill,MahlerRozema,HouZhu}, we are interested in analysing the performance of the proposed estimator $\hat{\rho}_n$ in terms of the \emph{maximum risk} with respect to the \emph{Bures distance} defined in equation \eqref{eq:maximumrisk}
and its asymptotic rescaled version 
\begin{equation*}
r_{\rm max}(\hat{\rho}) = 
\lim\sup_{n \rightarrow \infty}  \sup_{\rho} n \mathbb{E} \left[ D_B(\rho, \hat{\rho})^2 \right] =
\limsup_{n \rightarrow \infty} \, n R_{\rm max}(\hat{\rho}_n).
\end{equation*}
We will derive an upper bound for the latter risk, thereby demonstrating a $n^{-1}$ scaling for maximum risk over all states $\rho$. Since the maximum risk of any estimation procedure cannot scale faster than $n^{-1}$, this implies that the existence of a non-trivial scaling constant for the \emph{minimax risk} given by
$$
R_{\rm minmax} := \limsup_{n \rightarrow \infty} \,  \inf_{\hat{\rho}_n} n R_{\rm max}(\hat{\rho}_n).
$$
Finding the value of the minimax constant remains an open problem. We will come back to this problem in section \ref{sec:minimax} where it is shown that the minimax qubit estimation problem reduces to that of minimax estimation of a coin probability with respect to the square Hellinger distance risk.

The estimator we propose is constructed as follows. The first stage is a preliminary localisation step involving standard projective measurements of Pauli observables $\sigma_x, \sigma_y, \sigma_z$ on a fixed fraction $n_1$ of the total number of qubits $n$. An estimate of the direction vector $\mathbf{\tilde{r}}/ \vert \mathbf{\tilde{r}} \vert$ is constructed from the outcomes of these measurements. The following lemma shows that with high probability the estimated directional vector is within an angle of $O(n_1^{-1/2+\epsilon_1})$ of the true vector, where $\epsilon_1$ is a fixed (small) positive constant. 
\begin{lem}\label{lem:firsthoeff}
Let $X_i, Y_i, Z_i$ be the outcomes of measurements of $\sigma_x, \sigma_y,\sigma_z$ performed on independent qubits in state $\rho$ with Bloch vector $\mathbf{r}$, where $i=1,\dots , n_1/3$. Let $\mathbf{\tilde{r}}$ be the estimate of the Bloch vector, where each Bloch vector component is obtained by averaging the outcome results, e.g $\tilde{r}_x:= \frac{3}{n_1}\sum_i X_i $. Then we have that for $\epsilon_1 >0$, 
\begin{equation}
\mathbb{P}\left( \Vert \mathbf{r} - \mathbf{\tilde{r}} \Vert_2^2 > 6 n_1^{-1+2\epsilon_1} \right) \leq 6 \exp{\left( -\frac{2n_1^{2\epsilon_1}}{3} \right)}.
\end{equation}
\end{lem}
The proof of this lemma  follows directly from the Hoeffding's inequality applied to the binomial distribution corresponding to each of the Bloch vector components. The concentration inequality implies that when $\vert \mathbf{r} \vert$ is bounded away from zero, the magnitude of the angle $\Phi$ between the directional vectors $\mathbf{r}/ \vert \mathbf{r} \vert$ and $\mathbf{\tilde{r}}/ \vert \mathbf{\tilde{r}} \vert$ is of the order $O(n_1^{-1/2+\epsilon_1})$ with high probability.  

The second adaptive stage involves preforming measurements along this estimated direction. That is, projective measurements of the observable $\Xi := \vec{\boldsymbol{\sigma}}  \cdot {\mathbf{\tilde{r}}}/ \vert \mathbf{\tilde{r}} \vert$ are performed on the remaining $n_2:=n-n_1$ copies of the state. Let $k$ be the total number of $+1$ outcomes from these measurements. It is easy to see that $k$ is distributed binomially $B_{n_2,p}(k)$ with binomial parameter $p: = (1+\vert \mathbf{r} \vert \cos{\Phi})/2$. We estimate this parameter as $\hat{p}$ from the measurement outcomes using the `add-beta' estimator \cite{BraessDette,BraessSauer} defined as follows,
\begin{equation}\label{eq:addbeta1}
\hat{p}_{n_2} = 
\begin{cases}
\frac{1/2}{n_2+5/4},           & k = 0, \\
\frac{2}{n_2+7/4},              & k =1, \\
\frac{k+3/4}{n_2+3/2},  & k = 2, \ldots, n_2-2, \\
\frac{n_2-1/4}{n_2+7/4},        & k=n_2-1, \\
\frac{n_2+3/4}{n_2+5/4},       & k=n_2.
\end{cases}
\end{equation}
While this estimator is not in any sense optimal, it is known to be the minimax estimator for the Kulback-Leibler risk, which will be used below in deriving the upper bound for qubit tomography. 
 
The final estimate of the state puts together the estimate $\hat{p}$ and the estimated Bloch vector $\mathbf{\tilde{r}}$ as follows 
\begin{equation}
\hat{\rho}_n= \frac{1}{2}\left( I + \frac{2\hat{p}_{n_2}-1}{\vert \mathbf{\tilde{r}} \vert} \mathbf{\tilde{r}} \cdot \vec{\boldsymbol{\sigma}} \right). 
\end{equation}

It is easy to see that $\hat{p}_{n_2} = \hat{\lambda}$ by construction, where $\hat{\lambda}$ is the eigenvalue of the estimate $\hat{\rho}_n$. 

\subsection{An $n^{-1}$ scaling upper bound scaling }

We now look at deriving an upper bound for the Bures risk of this estimator, and demonstrate that the maximum over all states scales as $n^{-1}$. Recall that the Bures distance between the final estimate $\hat{\rho}_n$ and the true state $\rho$ is defined as 
$D_B(\rho, \hat{\rho}_n)^2 := 2\left[ 1- \sqrt{F(\rho, \hat{\rho}_n)} \right]$, where $F(\rho, \hat{\rho}_n)$ is the fidelity, expressed in terms of the Bloch vectors $\hat{\mathbf{r}}$ and $\mathbf{r}$ as
\begin{align}\label{eq:fidelityexpansion}
F(\rho, \hat{\rho}_n) &= \frac{1}{2} \left( 1+ \sqrt{1- \vert \mathbf{r} \vert^2} \sqrt{1-\vert \hat{\mathbf{r}} \vert^2} +  \mathbf{r} \cdot \hat{\mathbf{r}} \right) \notag \\
&=\frac{1}{2} \left( 1+ \sqrt{1- \vert \mathbf{r} \vert^2} \sqrt{1-\vert \hat{\mathbf{r}} \vert^2} +  \vert \mathbf{r} \vert \vert \hat{\mathbf{r}} \vert \cos{\Phi} \right),
\end{align} 
where $\Phi$ is  the angle between the Bloch vectors, or equivalently the angle between the vectors $\mathbf{\tilde{r}}$ and $\mathbf{r}$ by construction. From Lemma \ref{lem:firsthoeff}, this angle is known to be small and of the order $O(n^{-1/2 + \epsilon_1})$ with high probability. This implies that the cosine term in (\ref{eq:fidelityexpansion}) can be expanded to leading order in $\Phi$. In this case, the Bures distance is expressed as 
\begin{equation}\label{eq:buresexpansion}
D_B(\rho, \hat{\rho}_n)^2 = D_H(\boldsymbol{\lambda}, \hat{\boldsymbol{\lambda}})^2 + 
\frac{1}{4} \frac{(1-2\lambda)(1-2\hat{\lambda}_n)}{\sqrt{(1-\lambda)(1-\hat{\lambda}_n)} 
+ \sqrt{\lambda \hat{\lambda}_n}}  \Phi^2 +O(\Phi^4)
\end{equation}   
where $\lambda = (1- |{\bf r}|)/2, \hat{\lambda}_n = (1- |\hat{\bf r}_n|)/2$ are the smallest eigenvalues of $\rho$ and respectively 
$\hat{\rho}_n$, and 
$$
D_H(\boldsymbol{\lambda}, \hat{\boldsymbol{\lambda}})^2 := 
\left(\sqrt{\lambda}- \sqrt{\hat\lambda}\right)^2+ \left(\sqrt{1-\lambda}- \sqrt{1-\hat\lambda}\right)^2
$$ 
is the square Hellinger distance between the probability distributions $\boldsymbol{\lambda} = (\lambda , 1- \lambda)$ and $\hat{\boldsymbol{\lambda}} = (\hat\lambda , 1- \hat\lambda)$. The proof of this approximation can be found in Appendix \ref{appendix.2}.

Identifying $\hat{\boldsymbol{\lambda}} =(\hat{p}_{n_2},1-\hat{p}_{n_2})^{T}$, we upper bound the Bures distance as
\begin{align*}
D_B(\rho,\hat{\rho}_n)^2 &\leq D_H(\boldsymbol{\lambda},\boldsymbol{\hat{p}}_{n_2})^2 + \frac{1}{4} \Phi^2 + O(\Phi^4) \\
&\leq 2 D_H(\boldsymbol{\lambda} ,\boldsymbol{p})^2 + 2 D_H(\boldsymbol{p} , \boldsymbol{\hat{p}}_{n_2})^2 + \frac{1}{4} \Phi^2 + O(\Phi^4) 
\end{align*}
where the second inequality is established using the fact that the Hellinger distance satisfies the triangle inequality. 
%
Using the inequality $D_H^2 (\boldsymbol{\lambda}, {\bf p}) \leq 2 |\lambda - p |$ and $p = (1- \vert {\bf r} \vert \cos{\Phi})/2$ we further upper bound the risk as  
\begin{equation*}
D_B(\rho, \hat{\rho}_n)^2 \leq 2D_H(\boldsymbol{p}, \boldsymbol{\hat{p}}_{n_2})^2 + \left(\frac{1}{4} + |{\bf r}|\right) \Phi^2  + O(\Phi^4).
\end{equation*}
Taking expectation with respect to the measurement outcomes given the true state $\rho$, we have 
\begin{align*}
\mathbb{E} \left[ D_B(\rho, \hat{\rho}_n)^2 \right] &\leq 
2 \mathbb{E} \left[ 2D_H(\boldsymbol{p}, \boldsymbol{\hat{p}}_{n_2})^2 \right] + \frac{5}{4}\mathbb{E} \left[ \Phi^2  \right]  + O(n_1^{-2+4\epsilon_1})
\end{align*}
%
The maximum risk of the estimator is therefore bounded from above as 
\begin{eqnarray}
\sup_{\rho} \mathbb{E} \left[ D_B(\rho, \hat{\rho}_n)^2 \right] &\leq &
\sup_{\rho} 2\mathbb{E} \left[ D_H(\boldsymbol{p}, \boldsymbol{\hat{p}}_{n_2})^2 \right]  
 + O(n_1^{-1}) + O(n_1^{-2+4\epsilon_1}) \nonumber \\
&\leq & 
\sup_{\rho} 2\mathbb{E} \left[ D_{KL}(\boldsymbol{p}, \boldsymbol{\hat{p}}_{n_2}) \right] +  O(n_1^{-1}) + O(n_1^{-2+4\epsilon_1})  
\end{eqnarray} 
In the first inequality we upped bounded $\mathbb{E}[\Phi^2]$ as $O(n_1^{-1})$; this follows from the concentration inequality of 
Lemma \ref{lem:firsthoeff}. The second step employs the inequality between the squared Hellinger distance and the Kullback-Leibler (KL) distance, defined as $D_{KL}(\boldsymbol{\hat{p}},\boldsymbol{p}) := \hat{p} \log{\hat{p}/p} + (1-\hat{p}) \log{(1-\hat{p})/(1-p)}$. The reason for employing this inequality is that to the best of the authors' knowledge the asymptotic minimax optimal Hellinger risk of estimating the binomial parameter $p$ is not known in the literature. A discussion related to the difficulty in obtaining the asymptotic minimax Hellinger risk is left to section \ref{sec:minimax}. However, the minimax optimal rate for the KL loss function under the binomial distribution is known. The minimax optimal estimator is precisely the `add beta' estimator defined in (\ref{eq:addbeta1}) , and is known to achieve the asymptotic rate $\frac{1}{2n_2}(1+o(1))$ \cite{BraessDette,BraessSauer}.  Choosing $n_1$ to be constant fraction of the total number of samples $n$ establishes an overall rate of $O(1/n)$, for example by choosing $n_1 = n_2 = n/2$.

\section{Irreducible representations, collective measurements and local asymptotic normality}\label{sec:globalmeasurement}

In this section we propose a two stage estimator of the qubit state, which uses collective rather than separate measurements. 
The first stage, much like in the pervious section, is a preliminary localisation stage. However, this stage does not use a fixed fraction of the total number of copies of the state, but a vanishing fraction $\tilde{n}$ of the overall ensemble of identical qubit $n$. The second measurement stage involves performing joint measurements of the remaining $n-\tilde{n}$ copies of the state, and is based on the techniques of Local Asymptotic Normality (LAN) established in \cite{GutaJanssens,GutaKahn}. This section is structured as follows. In subsection \ref{sec:parameterisation} we describe the preliminary measurement stage, and define a choice of parameterisation of states. In subsection \ref{sec:jointmeasurement} we describe the joint measurement strategy. In subsection \ref{sec.lan} we provide a brief review of established LAN results and techniques which can be used to derive asymptotic estimation bounds. 
Minimax results for the proposed estimator are then established in section \ref{sec:bounds}.

\subsection{Preliminary localisation and parametrisation}\label{sec:parameterisation}

The first stage is a preliminary localisation step that involves performing standard projective measurements of the Pauli observables $\sigma_x, \sigma_y, \sigma_z$ on a vanishing fraction $\tilde{n}$ of the overall ensemble of $n$ identically prepared qubits. An estimate $\tilde{\rho}$ of the state is constructed from the outcomes of these measurements. The following lemma shows that \emph{with high probability} the true state $\rho$ lies within a ball of radius $O(n^{-1/2 + \epsilon_2})$ of this estimate $\tilde{\rho}$. This allows us to restrict our attention to a local neighbourhood of the preliminary estimator in the second stage of the estimation.
 
\begin{lem} \label{le:localisation}
Let $X_i, Y_i, Z_i$ be independent outcomes of measurements of $\sigma_x, \sigma_y,\sigma_z$ performed on independent qubits in state $\rho$ with Bloch vector ${\bf r}$, where $i=1,\dots , \tilde{n}/3$. Let $\tilde{\rho}$ be the estimator with Bloch vector 
$\tilde{\bf r}$ obtained by averaging the outcome results, e.g. $\tilde{r}_x:= \frac{3}{\tilde{n}}\sum_i X_i $.

In order to obtain a physical state, the final estimate of the state is constructed as 
\begin{equation}
\tilde{\rho} := \argmin_{\tau \in \mathcal{S}_2} \Vert \tau -  \left( \mathbb{1}+ \tilde{\boldsymbol{r}} \cdot \boldsymbol{\sigma} \right)/2 \Vert_1^2
\end{equation}
where the minimisation is over all the space of all $2 \times 2$ density matrices $\mathcal{S}_2$. For this estimator $\tilde{\rho}$, we have that for all $\epsilon_2 >0$,
\begin{equation}\label{eq:lemma1eq}
\mathbb{P}\left(\Vert \tilde{\rho} - \rho \Vert_1^2 > 3n^{2\epsilon_2-1}\right) \leq 6 \exp{\left(-\frac{2\tilde{n}n^{2\epsilon_2-1}}{3} \right)}, \ \ \ \ \ \ \mathrm{for~all~} \rho \in \mathcal{S}_2.
\end{equation} 
\end{lem}

The proof of this lemma follows from an application of Hoeffding's inequality, and can be found in Appendix \ref{appendix.1}. Setting $\tilde{n} = n^{1-\kappa}$, with $0< \kappa < 2\epsilon_2$, the probability of failure is exponentially small. The preliminary measurement stage therefore places the estimate $\tilde{\rho}$ in a local neighbourhood around the true state. Since $\Vert \tilde{\rho} -\rho \Vert_1^2 = \Vert \tilde{\boldsymbol{r}} - \boldsymbol{r} \Vert^2$, the angle between the two normalised Bloch vectors $\boldsymbol{r}/\vert \boldsymbol{r} \vert$ and $\tilde{\boldsymbol{r}}/\vert \tilde{\boldsymbol{r}} \vert$ is of the order $O(n^{-1/2+\epsilon_2})$ with high probability.


For technical reasons related to local asymptotic normality theory and the derivation of certain error bounds, the subsequent measurement depends on $\tilde{\rho}$, and we distinguish the following two cases. 

i) If $\tilde{\rho}$ is within a fixed but small ball of radius $\delta>0$ around the fully mixed state (i.e, $\vert \tilde{r} \vert \leq \delta$), the secondary measurement stage consists of the standard tomographic measurements in the $\sigma_i$, $i =x,y,z$ bases. For each $i$, measurements of $\sigma_i$ are performed on $(n-\tilde{n})/3$ identical copies of the state. The final estimate of the state $\hat{\rho}_n$ is constructed from the outcomes of these measurements, and is detailed in section \ref{sec:bounds}. 

ii) If $\tilde{\rho}$ is away from the fully mixed state, we can apply the tools of LAN. The remaining $n-\tilde{n}$ copies of the state available for the second stage are rotated such that the estimated Bloch vector $\tilde{\boldsymbol{r}}$ is pointing along the $z$-axis. From Lemma \ref{le:localisation}, the angle between the directional vectors $\boldsymbol{r}/\vert \boldsymbol{r} \vert$ and $\tilde{\boldsymbol{r}}/\vert \tilde{\boldsymbol{r}} \vert$ is known to be of the order $O(n^{-1/2+\epsilon_2})$ with high probability. This allows us to consider a restricted parametrisation of states which we describe now for an arbitrary but fixed state $\rho_0$ (which plays the role of $\tilde{\rho}$) with its Bloch vector along the $z$-axis
\begin{equation}\label{eq.rho0}
\rho_0 = 
\left(
\begin{array}{ccc}
1-\lambda_0 && 0 \\
0 && \lambda_0 
\end{array}
\right)
\end{equation}

with $ 0< \lambda_0 < 1/2$. We consider a parametrisation $\boldsymbol{\theta} \rightarrow \rho_{\boldsymbol{\theta}}$ of states obtained by small unitary rotations of $\rho_0$, and different choices of the eigenvalue. We choose the parameter vector $\boldsymbol{\theta} := (\lambda, \boldsymbol{w})$, where $\boldsymbol{w} = (u, v) \in \mathbb{R}^2$ corresponds to the small unitary rotations of the eigenvectors, and $\lambda$ is the smallest eigenvalue. That is, any state $\rho$ described by 
$\boldsymbol{\theta} = (\lambda,u,v)$  is of the form 
\begin{equation}
\rho_{\boldsymbol{\theta}} := U\left(\frac{\boldsymbol{w}}{\sqrt{n}}\right) 
\left(
\begin{array}{ccc}
1-\lambda && 0 \\
0 && \lambda 
\end{array}
\right)
U\left(\frac{\boldsymbol{w}}{\sqrt{n}}\right)^{*}
\end{equation}

where the unitary $U\left(\frac{\boldsymbol{w}}{\sqrt{n}}\right)$ is given by 
$$
U\left(\frac{\boldsymbol{w}}{\sqrt{n}}\right) := \exp\left(\frac{i}{\sqrt{n}} (u \sigma_x + v  \sigma_y)\right) = 
\left(
\begin{array}{ccc}
\cos \vert \boldsymbol{w} \vert /\sqrt{n} && -\exp(-i\varphi) \sin \vert \boldsymbol{w} \vert /\sqrt{n}  \\
\exp(i \varphi) \sin \vert \boldsymbol{w} \vert/\sqrt{n}   && \cos \vert \boldsymbol{w} \vert  /\sqrt{n} 
\end{array}
\right)
$$
with $\varphi ={\rm Arg} (-v + i u)$. Note that in this parametrisation we have 
$\rho_0 = \rho_{\boldsymbol{\theta}_0}$ with $\boldsymbol{\theta}_0 = (\lambda_0,0,0)$. 
The aim of the second measurement stage is then to estimate the unknown parameter vector $\boldsymbol{\theta} = (\lambda, u, v) = (\lambda, \boldsymbol{w})$ corresponding to the true state $\rho$.

\subsection{The `which block' measurement stage}\label{sec:jointmeasurement}

The second measurement stage involves a joint measurement on the $n-\tilde{n}$ remaining copies of the state. We therefore consider the joint states $\rho^{\boldsymbol{\theta}}_n := \rho_{\boldsymbol{\theta}}^{\otimes n}$ on $n$ identical qubits, with the parametrisation around the preliminary estimator $\rho_0 = \tilde{\rho}$ described above. It is known that the states $\rho^{\boldsymbol{\theta}}_n$ have a block-diagonal form with respect to the decomposition of the underlying space $(\mathbb{C}^2)^{\otimes n}$ in irreducible representations  of the groups $SU(2)$ and $S(n)$ \cite{GutaJanssens,BaganBallester,CiracEkert}. The representation $\pi_n$ of $SU(2)$ is given by $\pi^{(n)}(u) = u^{\otimes n}$ for any $u \in SU(2)$, and the representation $\tilde{\pi}_n$ of the symmetric group $S(n)$ is given by the permutation of factors 
\begin{equation}
\tilde{\pi}^{(n)}(\tau) : v_1 \otimes \ldots \otimes v_n \rightarrow v_{\tau^{-1}(1)} \otimes \ldots \otimes v_{\tau^{-1}(n)}, \ \ \ \ \ \ \ \tau \in S(n) 
\end{equation}
According to Weil's Theorem, the following decomposition holds
\begin{equation}
(\mathbb{C}^2)^{\otimes n} = \bigoplus_{j=0, 1/2}^{n/2} \mathcal{H}_j \otimes \mathcal{H}^{j}_n
\end{equation}
where the lower limit in the direct sum is $0$ for even $n$ and $1/2$ for odd $n$. The two group representations decompose into direct sums of irreducible representations as $\pi^{(n)} (u)= \oplus_j \pi_j (u)\otimes \mathbb{1}$ and $\tilde{\pi}^{(n)} (\tau)= \oplus_j \mathbb{1} \otimes \tilde{\pi}_j (\tau)$ where $ \pi_j $ is the irreducible representation of $SU(2)$ with total angular momentum $J^2 = j(j+1)$ which acts on $\mathcal{H}_j \cong \mathbb{C}^{2j+1}$, and  $\tilde{\pi}_j$  is the irreducible representation of the symmetric group $S(n)$ acting on 
$\mathcal{H}^j \cong \mathbb{C}^{n_j}$ with 
\begin{equation*}
n_j = \binom{n}{n/2-j} - \binom{n}{n/2-j-1}, \ \ \ \ \ \ \ \ \ \ j \neq n/2
\end{equation*}

and $n_{j=n/2} = 1$. The density matrix $\rho^{\boldsymbol{\theta}}_n$ is invariant under permutations and can be decomposed as 
\begin{equation*}
\rho^{\boldsymbol{\theta}}_n = \bigoplus_{j=0,1/2}^{n/2} p_{n,\lambda}(j)\rho^{\boldsymbol{w}}_{j,n} \otimes \frac{\boldsymbol{1}}{n_j}
\end{equation*}
where the probability distribution $p_{n,\lambda}(j)$ is given by \cite{GutaJanssens,GutaKahn,BaganBallester} 
\begin{equation}\label{eq.distribution.blocks}
p_{n,\lambda}(j):= \frac{n_j}{1-2\lambda} \lambda^{n/2-j} (1-\lambda)^{n/2+j+1} (1-p^{2j+1})
\end{equation}
with $p = \frac{\lambda}{1-\lambda}$. The above distribution can be written in the form 
\begin{equation}\label{eq:blockprob}
p_{n,\lambda}(j) := B_{n,\lambda}(n/2-j) \times K(j,n,\lambda), 
\end{equation}
where $B_{n,\lambda}(k) = \binom{n}{k} \lambda^k (1-\lambda)^{n-k}$ is the binomial distribution and the term $K(j,n,\lambda)$ is given by
\begin{equation*}
K(j,n, \lambda) := (1-p^{2j+1}) \frac{n+(2(j-j_n)+1)/(1-2\lambda)}{n+(j-j_n+1)/(1-\lambda)}, \ \ \ \ \ j_n := n(1/2-\lambda).
\end{equation*}
The binomially distributed variable $n/2-j$ concentrates around its mean value of $n\lambda$ with high probability 
\begin{equation*}
\mathbb{P}\left[ n\lambda - n^{1/2+\epsilon_3} \leq n/2-j \leq n\lambda + n^{1/2+\epsilon_3} \right] \geq 1- 2 \exp(-2n^{2\epsilon_3}),
\end{equation*}

where $\epsilon_3 >0$ is an arbitrary constant. This follows from a straightforward application of  Hoeffding's inequality (\ref{eq:hoeffding}) to the binomial distribution. The mass of the distribution $B_{n,\lambda}(n/2-j)$ therefore concentrates over values of $j$ in the interval
\begin{equation}\label{eq:jinterval}
\mathcal{J}_n := \{ j \vert \ j_n - n^{1/2+\epsilon_3} \leq j \leq j_n + n^{1/2+\epsilon_3} \}.
\end{equation}

For all $j \in \mathcal{J}_n$, the factor $K(j,n,\lambda) = 1+O(n^{-1/2+\epsilon_3})$ provided that $\lambda$ is bounded away from $1/2$, which is one of the reasons we chose to treat the two cases above separately. Additionally we note that the factor $K(j,n,\lambda)$ remains bounded over all values of $j$ as long as $\lambda < 1/2$. From the concentration of the binomial distribution over values of $j \in \mathcal{J}_n$, and the value of $K(j,n,\lambda)$ in this interval it follows that
\begin{equation}
p_{n,\lambda}(\mathcal{J}_n) = 
1 - O(n^{-1/2+\epsilon_3}).
\end{equation}
A ``which block" measurement corresponds to an output of a particular value of $j$ from the distribution (\ref{eq:blockprob}), and an associated posterior state $\rho^{\boldsymbol{w}}_{j,n}$. This value of $j$ lies in the set $\mathcal{J}_n$ with high probability. The eigenvalue parameter $\lambda$ is estimated form this value of $j$. As in the case of the local adaptive estimator in section \ref{sec:localmeasurement}, in order to derive a minimax upper bound, we shall define $\hat{\lambda}$ as the `add-beta' estimator, identifying $n/2-j$ with $k$ in (\ref{eq:addbeta1}). However a discussion regarding this choice for the estimator $\hat{\lambda}$ is discussed later in section \ref{sec:minimax}. We note that a possible physical implementation of such a measurement is detailed in \cite{GutaJanssens}, and involves coupling the joint states to different bosonic field and performing a homodyne measurement.

Information about the `rotation' parameters are contained in the block state $\rho^{\boldsymbol{w}}_{j,n}$. These parameters are estimated using established LAN results \cite{GutaJanssens} which we recall in section \ref{sec.lan} below. 
 
 \subsection{Local asymptotic normality} \label{sec.lan}
 
 The block state $\rho^{\boldsymbol{w}}_{j,n}$ encodes information about the rotation parameters $\boldsymbol{w} =(u,v)$. The optimal estimation strategy for these parameters has been established using results about the LAN of qubit states \cite{GutaJanssens}. 
This shows that for large $n$, the block states $\rho^{\boldsymbol{w}}_{j,n}$ approach a Gaussian state $\phi^{\boldsymbol{w}}$ of a one-mode continuous variables system \emph{uniformly} over all $j \in \mathcal{J}_n$ and $\Vert \boldsymbol{w} \Vert \leq n^{\eta}$. The rotation parameters $(u,v)$ are encoded linearly into the mean of the Gaussian state $\phi^{\boldsymbol{w}}$. So the problem of the optimal estimation of these parameters for the block state can be translated into one of estimating the displacement of $\phi^{\boldsymbol{w}}$. These ideas have been treated in detail in \cite{GutaJanssens}, and we only include a brief overview here.  
The block states $\rho^{\boldsymbol{w}}_{j,n}$ depend on the parameters $(u,v)$ in the following way 
\begin{equation}\label{eq:blockstate}
\rho^{\boldsymbol{w}}_{j,n} = U_j \left( \frac{\boldsymbol{w}}{\sqrt{n}} \right) \rho^{\boldsymbol{0}}_{j,n} U_j \left( \frac{\boldsymbol{w}}{\sqrt{n}} \right)^{*}
\end{equation}
where the unitaries are defined as $U_j (\boldsymbol{w}) := \exp(i (u J_{j,x} + v J_{j,y})$, with 
$J_{j,l}$ being the generators of rotations in the irreducible representation $\pi_j$ of SU(2). The state $ \rho^{\boldsymbol{0}}_{j,n}$ is expressed as 
\begin{equation}
\rho^{\boldsymbol{0}}_{j,n} = \frac{1-p}{1-p^{2j+1}} \sum_{m=-j}^{j} p^{j-m} \vert j, m \rangle \langle j,m \vert 
\end{equation}
with $p= \lambda/(1-\lambda)$ as before. The set $\{ \vert j, m \rangle : m=-j, \ldots, j \}$ is an orthonormal basis on $\mathcal{H}_j$ such that $J_{j,z} \vert j,m \rangle = m \vert j,m \rangle$. It has been demonstrated \cite{GutaJanssens} that the family of states $\mathcal{F}_n := \{ \rho^{\boldsymbol{w}}_{j,n} , \Vert \boldsymbol{w} \Vert \leq n^{\eta}, j\in \mathcal{J}_n \}$ is asymptotically Gaussian. This mean that as $n \rightarrow \infty$ the family of states $\rho^{\boldsymbol{w}}_{j,n}$ ``converges'' to a family of Gaussian states $\phi^{\boldsymbol{w}}$ of a one-mode continuous variables system, for all $j \in \mathcal{J}_n$ and $\Vert \boldsymbol{w} \Vert \leq n^{\eta}$. In order to make this convergence more precise, we let 
\begin{equation}
\phi^{\boldsymbol{0}} := (1-p) \sum_{k=0} p^{k} \vert k \rangle \langle k \vert 
\end{equation}
be a centred Gaussian state of a one mode continuous variables system, with $\{\vert k \rangle : k\geq 0\} $ denoting the Fock basis. 
The states $\phi^{\boldsymbol{w}}$ are defined as 
\begin{equation}\label{eq:gaussianstate}
\phi^{\boldsymbol{w}} := D(\sqrt{1-2\lambda} \alpha_{\boldsymbol{w}}) \phi^{\boldsymbol{0}} D(-\sqrt{1-2\lambda} \alpha_{\boldsymbol{w}})
\end{equation}
where $\alpha_{\boldsymbol{w}} = -v + i u\in \mathbb{C}$. The operator $D(\alpha) := \exp(\alpha a^{*} - \overline{\alpha} a)$ is the displacement operator that for every $\alpha \in \mathbb{C}$ maps the vacuum vector $\vert 0 \rangle$ to the coherent state $\vert \alpha \rangle$, with $a^{*}, a$ being the creation and annihilation operators satisfying $[ a, a^{*} ] =1$. The convergence of the block state $\rho^{\boldsymbol{w}}_{n,j}$ to the Gaussian state $\phi^{\boldsymbol{w}}$ is formalised in the following Theorem. 

\begin{theorem}\label{le:LAN}
 Let $V_j : \mathcal{H}_j \rightarrow L^2(\mathbb{R})$ be the isometry 
\begin{equation}
V_j : \vert j,m \rangle \rightarrow \vert j-m \rangle 
\end{equation}
that maps the orthonormal basis of $\mathcal{H}_j$ into the Fock basis of $L^2(\mathbb{R})$. 
Then for the family of block states $\rho^{\boldsymbol{w}}_{n,j}$  defined by (\ref{eq:blockstate}), and the family of Gaussian states $\phi^{\boldsymbol{w}}$ defined by (\ref{eq:gaussianstate}), the following convergence holds for any $0 \leq \eta \leq 1/6$ and $0< \epsilon_3 < 1/2$
\begin{equation}
\sup_{\Vert \boldsymbol{w} \Vert \leq n^{\eta}} \max_{j \in \mathcal{J}_n} \Vert V_j \rho^{\boldsymbol{w}}_{j,n} V_j^{*} - \phi^{\boldsymbol{w}} \Vert_1 = O(n^{-1/4+\eta+\epsilon_3}) 
\end{equation}
over the set $\mathcal{J}_n = \{ j \ \vert \ j_n - n^{1/2+\epsilon_3} \leq j \leq j_n + n^{1/2+\epsilon_3} \}$. The convergence is uniform over 
$\lambda \geq 1/2 (1+\delta)$ for an arbitrary fixed $\delta>0$.
\end{theorem}
The interpretation is that the block state $\rho^{\boldsymbol{w}}_{j,n}$ can be mapped by means of physical transformations (in this case an isometric embedding) into the Gaussian state $\phi^{\boldsymbol{w}}$ with vanishing norm-one error, uniformly over the unknown parameter ${\boldsymbol{w}}$ and over the block index $j$. A possible physical implementation is detailed in \cite{GutaJanssens}; the ensemble of qubits is coupled with a Bosonic field such that the state is transferred to the field after some time. 

In order to estimate the rotation parameters $\hat{\boldsymbol{w}}= (\hat{u},\hat{v})$, one first maps the qubit state via the isometry $V_j$, and then performs a heterodyne measurement, which is optimal for estimating displacement. In the next section we discuss the Bures risk of the estimation procedure described above.

\section{Upper and lower bounds for collective measurements}\label{sec:bounds} 

The overall measurement procedure we propose can be briefly summarised as consisting of a preliminary localisation stage, where a vanishing number $\tilde{n}$ of copies of the state is used to localise the state $\rho$. This estimate $\tilde{\rho}$ informs the choice of measurements in the second stage. When $\tilde{\rho}$ is within a fixed ball of radius $\delta>0$ around the fully mixed state,  standard tomographic measurements are performed on the remaining copies of the state. However, if $\tilde{\rho}$ lies outside this ball, the measurements performed in second stage uses techniques based on the principle of LAN to estimate the parameter vector $\boldsymbol{\theta} =(\lambda, u, v)$. In this section we look at the Bures risk of the measurement estimator pair 
$
R(\rho, \hat{\rho}_n) := \mathbb{E} \left[D_B(\rho, \hat{\rho}_n)^2 \right].
$ 
As we will show below, the risk of a good estimators scales as $1/n$, and we would like to find asymptotic upper and lower bounds for the rescaled maximum risk  
$$
R_{max} (\hat{\rho}) =  \limsup_{n\to\infty} ~\sup_{\rho \in \mathcal{S}_2}~  n R(\rho, \hat{\rho}_n). 
$$

\subsection{The upper bound} \label{sec:upper.bound.collective}


We now make concrete our final estimator for the state $\rho$. The first stage involves using a vanishing number of copies $\tilde{n}:=n^{1-\kappa}$ (with $\kappa>0$) to get a rough estimate $\tilde{\rho}$. This estimate informs the second measurement stage. The subsequent measurement stage differs depending on whether the state is estimated to be close to the fully mixed state. 

If the estimate $\tilde{\rho}$ lies in a small ball of radius $\delta>0$ around the fully mixed state, then measurements in the standard $\sigma_i$, $i = x,y,z$ basis are performed on $(n-\tilde{n})/3$ copies of the state. The outcomes of each measurement $\pm 1$, and the associated probabilities are $p(\pm1 \vert \sigma_i) = p_i(\pm1) := \rm{Tr}(\rho P_i^{\pm1})$, where the projectors $P^{\pm1}_i$ are defined via $\sigma_i = P^{+1}_i - P^{-1}_i$.   The total number $n_i$ of $+1$ outcomes obtained by $n/3$ measurements of $\sigma_i$ is binomially distributed $B_{n/3,p_i(+1)}(n_i)$. The final estimate of the state is constructed as the maximum likelihood (ML) estimate from these measurement outcomes 
\begin{equation}\label{eq:case1}
\hat{\rho}_n = \arg\max_{\tau \in \mathcal{S}_2} \sum_{i=x,y,z} n_i \log{{\rm Tr}(\tau P^{+1}_i)} + (n/3-n_i) \log{{\rm Tr}(\tau P^{-1}_i)}
\end{equation}
where the maximisation is over the space of all $2 \times 2$ density matrices $\mathcal{S}_2$. 

On the other hand, if the preliminary estimate $\tilde{\rho}$ lies away from the fully mixed state, we perform the following measurements to estimate the parameter vector $\boldsymbol{\theta} = (\lambda, u, v)$.  A `which block' measurement outputs a value of $j$ from which the eigenvalue $\lambda$ is estimated, cf. section \ref{sec:jointmeasurement}. Similarly to the separate measurements strategy, we consider the following `add-beta' estimator for the eigenvalue $\lambda$ \cite{BraessDette,BraessSauer}
\begin{equation}\label{eq:addbetaKL}
\hat{\lambda}_n = 
\begin{cases}
\frac{1/2}{n+5/4},           & \frac{n}{2}-j = 0, \\
\frac{2}{n+7/4},              & \frac{n}{2}-j =1, \\
\frac{n/2-j+3/4}{n+3/2},  & \frac{n}{2}-j = 2, \ldots, n-2, \\
\frac{n-1/4}{n+7/4},        & \frac{n}{2}-j=n-1, \\
\frac{n+3/4}{n+5/4},       & \frac{n}{2}-j=n
\end{cases}
\end{equation}

The range of possible values of $j$ is $[0,n/2]$, and therefore only some of the rules of the estimator described above are used. However, we describe the estimator over the range $[0,n]$ as this will be used shortly to upper bound the minimax risk.

Conditional on $j$, we are left with the block state $\rho_{j,n}^{\boldsymbol{w}}$. 
Using Theorem \ref{le:LAN} we can isometrically map this state onto the Fock space, close to the Gaussian state 
$\phi^{\boldsymbol{w}}$. In order to estimate the displacement parameter $\boldsymbol{w}$ we perform a heterodyne measurement with outcome $\hat{\boldsymbol{w}}_n$. The final estimate of our the state $\hat{\rho}_n$ is constructed from the estimated parameter vector 
$\hat{\boldsymbol{\theta}}_n = (\hat{\lambda}_n, \hat{\boldsymbol{w}}_n )$ as
\begin{equation}\label{eq:case2}
\hat{\rho}_n = U\left(\frac{\hat{\boldsymbol{w}}_n}{\sqrt{n}}\right)  
\left(
\begin{array}{ccc}
1-\hat{\lambda}_n && 0 \\
0 && \hat{\lambda}_n 
\end{array}
\right)
U\left(\frac{\hat{\boldsymbol{w}}_n}{\sqrt{n}}\right)^{*}.
\end{equation}

We now state precisely the an upper bound for the minimax risk of the Bures distance for the measurement strategy described above. 
\begin{theorem}\label{th:upperbound}
Let  $\hat{\rho}_n$ be the estimator described above. The asymptotic rescaled maximum risk is bounded from above as 
\begin{equation}
\limsup \limits_{n\rightarrow \infty} \  \sup_{\rho}\  nR(\rho,\hat{\rho}_n) \leq \frac3 2 .
\end{equation} 
\end{theorem}  

The proof of this theorem is detailed in the appendix (\ref{proofofthm}), and here we only provide an outline for the arguments employed. The choice of measurements in the second stage depend on whether the preliminary estimate $\tilde{\rho}$ lies inside or outside a small ball of radius $\delta>0$ around the fully mixed state. In keeping with this, let us therefore denote $\hat{\rho}_n^1$ as the estimator (\ref{eq:case1}) chosen when $\vert \tilde{r} \vert \leq \delta$, and let $\hat{\rho}_n^2$ be the LAN based estimator (\ref{eq:case2}), when $\vert \tilde{r} \vert > \delta$. The minimax risk can then be bounded from above as follows 
\begin{equation}\label{maxrisk}
\begin{split}
\limsup \limits_{n\rightarrow \infty} \  \sup_{\rho}\  nR(\rho,\hat{\rho}_n) \leq  \max \bigg\{ \limsup \limits_{n\rightarrow \infty} \sup_{\rho \in B_1} & n\mathbb{E}\left[ D_B(\rho,\hat{\rho}^1_n)^2 \big\vert \ \vert \tilde{r} \vert \leq \delta  \right], \\
&\limsup \limits_{n\rightarrow \infty} \sup_{\rho \not\in B_2} n\mathbb{E}\left[ D_B(\rho,\hat{\rho}^2_n)^2 \big\vert \  \vert \tilde{r} \vert > \delta \right] \bigg\} 
\end{split}
\end{equation}

where $B_1$ and $B_2$ are balls of radius $\delta + n^{-1/2 + \epsilon_2}$ and  $\delta - n^{-1/2 + \epsilon_2}$ respectively. The two terms are evaluated explicitly in section (\ref{proofofthm}) of the appendix. The term corresponding to the estimator $\hat{\rho}_n^1$ is straightforward to evaluate as the Bures distance is locally quadratic for states in $B_1$. From this quadratic expansion and the efficiency of the maximum likelihood estimator in the asymptotic regime, the risk can be expressed as 
\begin{equation}
\mathbb{E} \left[ D_B(\rho, \hat{\rho}^1_n)^2 \right] \approx \frac{1}{n}{\rm Tr} \left(I(\rho)^{-1}G\right)
\end{equation}

where $G$ is the weight matrix reconstructing the quadratic approximation of the Bures distance, and $I$ is the Fisher information matrix. From the explicit form of $G$ and $I$, the asymptotic risk can be bounded as
 \begin{equation}\label{eq:case1r} 
\limsup \limits_{n\rightarrow \infty} \  \sup_{\rho \in B_1} n\mathbb{E} \left[ D_B(\rho,\hat{\rho}^1_n)^2 \right]
\leq \frac{3}{4} \left( 1+ \frac{\delta}{1-\delta} \right). 
\end{equation} 

The other term in (\ref{maxrisk}) corresponding to the estimator $\hat{\rho}_n^2$ uses the local parameterisation of states $\boldsymbol{\theta} = (\lambda, \boldsymbol{w})$, and the approximation of the Bures distance used in section \ref{sec:localmeasurement}, and detailed in the appendix (\ref{appendix.2}). We therefore get that the risk can be bounded as
\begin{align}
\mathbb{E} \left[ D_B(\rho,\hat{\rho}^2_n)^2  \right] &\leq \mathbb{E} \left[ D_H(\boldsymbol{\lambda}, \hat{\boldsymbol{\lambda}}_n)^2  + \frac{1}{4}  \Phi^2 \right] +O(\Phi^4)\\
&\leq \mathbb{E} \left[  D_H(\boldsymbol{\lambda}, \hat{\boldsymbol{\lambda}}_n)^2 \right] +\frac{1}{n}\mathbb{E} \left[  (u-\hat{u}_n)^2 + (v - \hat{v}_n)^2 \right] +O(n^{-2})
\end{align}

The term corresponding to the rotation parameters has been evaluated in \cite{GutaJanssens} using LAN based techniques. Since LAN holds in the limit of large $n$, the problem of estimating the rotation parameters is translated to one of determining the displacement of a Gaussian state $\phi^{\boldsymbol{w}}$. The heterodyne measurement is known to be the optimal measurement in this case \cite{GutaJanssens,GutaKahn}. The first term corresponding to the Hellinger risk is bounded from above by the Kullback-Leibler (KL) risk of estimating a binomial parameter $\lambda$ from outcomes $k$ distributed as $B_{n,\lambda}(k)$.  The `add-beta' estimator is known to be minimax optimal in this case, and its rate is known in the literature. Together with the LAN results for the rotation parameters, we bound risk as 
 
\begin{equation}
\limsup \limits_{n \rightarrow \infty} \sup_{\rho \not\in B_2} n R(\rho, \hat{\rho}^2_n) \leq \limsup \limits_{n \rightarrow \infty} \sup_{\lambda} n \mathbb{E}_{\text{Binom}} \left[ D_{KL}(\boldsymbol{\lambda},\hat{\boldsymbol{\lambda}}_n) \right]  + 1 
\leq  \frac{3}{2}.
 \end{equation}
 
 Comparing this bound with (\ref{eq:case1r}), we arrive at the stated upper bound of $3/2$ in Theorem \ref{th:upperbound} provided $\delta < 1/2$. 

\subsection{The lower bound}\label{sec.lower.bound}
In this section we derive a lower bound on the asymptotic rescaled risk with respect to the Bures distance. 
The key idea is to restrict the attention to a smaller state space region where the state is ``hardest'' to estimate, and evaluate the minimax risk over this region, thus obtaining a lower bound for the overall minimax risk.

Let us consider that the true state $\rho$ lies in a local neighbourhood of size $n^{-1/2 + \epsilon}$ around an arbitrary but fixed state $\rho_0$ as defined in equation \eqref{eq.rho0}, whose smallest eigenvalue satisfies $0<\lambda_0<1/2$.
For any estimation procedure $\hat{\rho}_n$ we have the lower bound for the maximum risk
\begin{eqnarray}
 \limsup \limits_{n \rightarrow \infty} \sup_{\rho\in \mathcal{S}_2} n R(\rho, \hat{\rho}_n)
&\geq& 
\limsup \limits_{n \rightarrow \infty} \sup_{\Vert \rho - \rho_0 \Vert_1 \leq n^{-1/2+ \epsilon}} n R(\rho, \hat{\rho}_n) \nonumber \\
&\geq &
\limsup_{n \rightarrow \infty}  ~\inf_{\hat{\rho}_n} ~ \sup_{\Vert \rho - \rho_0 \Vert_1 \leq n^{-1/2+ \epsilon}}  n R(\rho, \hat{\rho}_n)\nonumber \\
&:= & R_{minmax}(\rho_0) 
\end{eqnarray} 
where the right side is the \emph{local minimax} risk at $\rho_0$.

Since the state $\rho_0$ is taken to be away from the boundary of the Bloch sphere, we can parametrise its local neighbourhood using the local parameter $\boldsymbol{\theta}= (h, {\bf u})$
\begin{equation}\label{eq.parametrisation.lower.bound}
\rho = \rho_{\boldsymbol{\theta}} := U\left(\frac{\boldsymbol{w}}{\sqrt{n}}\right)  
\left(
\begin{array}{ccc}
1-\lambda_0 -h/\sqrt{n} && 0 \\
0 && \lambda_0 + h/\sqrt{n} 
\end{array}
\right)
U\left(\frac{\boldsymbol{w}}{\sqrt{n}}\right)^{*}.
\end{equation}
The Bures distance is locally quadratic 
\begin{equation}\label{eq:buresquadratic}
D_B(\rho_{\boldsymbol{\theta}}, \rho_{\boldsymbol{\theta}^\prime})^2 = \frac{1}{n}(\boldsymbol{\theta}-\boldsymbol{\theta}^\prime)^T \Gamma_0 (\boldsymbol{\theta}- \boldsymbol{\theta}^\prime) + O(n^{-3/2})
\end{equation}
where $\Gamma_0$ is the weight matrix 
\begin{equation}
\Gamma_0 = 
\begin{pmatrix}
\frac{1}{4\lambda_0(1-\lambda_0)} & 0 & 0 \\
0 & (1-2\lambda_0)^2 & 0 \\
0 & 0 & (1-2\lambda_0)^2
\end{pmatrix}.
\end{equation}

In this case we can apply the LAN theory \cite{GutaJanssens} to obtain the local minimax risk for Bures distance. 
The upshot of the theory is that the classical statistical model given by the distribution over blocks (cf. equation \eqref{eq.distribution.blocks}) can be approximated by a one-dimensional Gaussian model $N(h, v_0)$ with fixed variance 
$v_0=\lambda_0 (1-\lambda_0) $ and mean equal to the unknown local parameter $h$. Additionally, the quantum statistical model described by the quantum state of the irreducible block can be approximated by a quantum Gaussian shift model (independent of the classical one), as described in Theorem \ref{le:LAN}. The optimal measurement here is the heterodyne, and after rescaling by a constant factor we obtain the unbiased estimator $\hat{\bf u}$ which has a two-dimensional Gaussian distribution 
$\hat{\bf u}\sim N({\bf u} ,w_0 \cdot I_2 )$ with $w_0 = (1-\lambda_0)/(2 (1- 2\lambda_0)^2) $. The local minimax risk is the sum of the contribution from the classical and respectively the quantum part of the Gaussian model, weighted with the matrix $\Gamma_0$
\begin{eqnarray}
R_{minmax}(\rho_0) 
&=&
\Gamma_{00} \mathbb{E}[(\hat{h}-h)^2] + \Gamma_{11} \mathbb{E}[(\hat{u}-u)^2] + 
\Gamma_{22} \mathbb{E}[(\hat{v}-v)^2] \nonumber \\  
&=&    
 \frac{1}{4\lambda_0 (1-\lambda_0)} v_0 + 2 (1-2\lambda_0)^2 w_0 = \frac{1}{4}+ (1-\lambda_0) = 
\frac{5}{4} - \lambda_0.
 \label{eq:localminmax}
\end{eqnarray}
As the state $\rho_0$ defining the local neighbourhood is chosen arbitrarily, we see that the right side of the above equation achieves its maximum as $\lambda_0 \rightarrow 0$, and we therefore get the asymptotic lower bound for the rescaled maximum risk of any estimator. 
$$
 \limsup \limits_{n \rightarrow \infty} \sup_{\rho\in \mathcal{S}_2} n R(\rho, \hat{\rho}_n) \geq \frac{5}{4}.
$$
As expected, the above lower bound is smaller than the $3/2$ upper bound derived in section \ref{sec:upper.bound.collective}.



 \section{The minimax optimal estimator}\label{sec:minimax}
 
 In deriving the minimax bounds for both the proposed estimators, the key observation was that the Bures risk decomposes locally into contributions from the Hellinger risk of estimating the eigenvalue parameter $\lambda$ and a quadratic risk corresponding to the estimation of the rotation parameters (see (\ref{eq:buresexpansion}) and appendix \ref{appendix.2}). The Hellinger risk was then bounded from above by the Kullback-Leibler (KL) risk of estimating the binomial parameter. The estimator of the binomial parameter achieving the minimax rate for the KL risk is known to be `add-beta' estimator, and both the local and global estimators for the state $\rho$ proposed using this estimator for the eigenvalue parameter (\ref{eq:addbetaKL},\ref{eq:addbeta1}).

The reason why we were not able to prescribe an asymptotically minimax estimator is that we could not devise a minimax estimator for the binomial parameter $\lambda$, with respect to the Hellinger distance. The following proposition follows immediately from the asymptotic analysis of section \ref{sec:bounds} and shows that the original optimal state estimation problem reduces to the `classical' one of estimating the binomial parameter $\lambda$.

\begin{prop}
Let $\hat{\lambda}_{\text{opt}}$ be an asymptotically minimax estimator of the binomial parameter $\lambda$ under the Hellinger loss function. The estimators defined by replacing the `add-beta' estimators for $\lambda$ in equation (\ref{eq:addbetaKL}) 
with $\hat{\lambda}_{\text{opt}}$ will then be asymptotically minimax optimal for qubit states. 
\end{prop} 

Although we were not able to devise a minimax estimator under the Hellinger distance, we would like to make some comments on this  problem, emphasising that it is crucial to study what happens at the boundary when $\lambda\approx 0$. Indeed, for values of $\lambda$ away from this boundary, the local asymptotic minimax rate is easily derived as the squared Hellinger distance square is locally quadratic and the classical asymptotic efficiency theory \cite{YoungSmith} applies. The standard estimator $\hat{\lambda} = k/n$ is a natural first choice as it is unbiased and achieves the Cramer Rao lower bound with variance $\text{Var}(\hat{\lambda}) = (nI)^{-1}$, where $I = \frac{1}{\lambda(1-\lambda)}$ is the Fisher information. In the region where $\lambda>0$, using a locally quadratic approximation for the Hellinger risk, we have
\begin{equation*}
\mathbb{E}_{\text{Binom}} \left[ D_H(\boldsymbol{\lambda},\hat{\boldsymbol{\lambda}})^2 \right] = \frac{1}{4 \lambda(1-\lambda)} \text{Var}(\hat{\lambda}) + 
o(n^{-1})
= \frac{1}{4n} + o(n^{-1}) .
\end{equation*}

This holds for every \emph{fixed} $\lambda \in (0, 1/2]$, and gives the same rate as the one in the lower bound (\ref{eq:localminmax}). However the convergence is not uniform over $\lambda$ close to the zero, which affects the constant in the asymptotic maximum risk. To see this, consider the case when $\lambda$ is $n$ dependent such that $n \lambda \rightarrow \mu$, with $\mu>0$ being a fixed constant. The Hellinger risk is given by 
\begin{equation}\label{eq:tmp1}
\mathbb{E}_{\text{Binom}} \left[ D_H(\boldsymbol{\lambda},\hat{\boldsymbol{\lambda}})^2 \right] = \mathbb{E}_{\text{Binom}} \left[ \left(\hat{\lambda}^{1/2} -\lambda^{1/2} \right)^2\right] + \mathbb{E}_{\text{Binom}}\left[\left( (1-\hat{\lambda})^{1/2} -(1-\lambda)^{1/2} \right)^2 \right] .
\end{equation}

The second term in the above equation is bounded as 
\begin{equation}\label{eq:tmp2}
\mathbb{E}_{\text{Binom}}\left[ \left( (1-\hat{\lambda})^{1/2} -(1-\lambda)^{1/2} \right)^2\right] \leq \mathbb{E}_{\text{Binom}}\left[ (\lambda - \hat{\lambda} )^2 \right]/(1-\lambda) = \mu/n^2.
\end{equation} 
Substituting (\ref{eq:tmp2}) in (\ref{eq:tmp1}), we have 
\begin{align}
\mathbb{E}_{\text{Binom}} \left[ D_H(\boldsymbol{\lambda},\hat{\boldsymbol{\lambda}})^2 \right] &=\mathbb{E}_{\text{Binom}}\left[ (\hat{\lambda}^{1/2} -\lambda^{1/2})^2 \right] + O(n^{-2}) \\ 
&= \frac{1}{n}\mathbb{E}_{\text{Po}(\mu)}\left[ \left( K^{1/2} -\mu^{1/2}\right)^2 \right]+ O(n^{-2})
\end{align}

  \begin{figure}[t]
 \centering
  \includegraphics[scale=0.55]{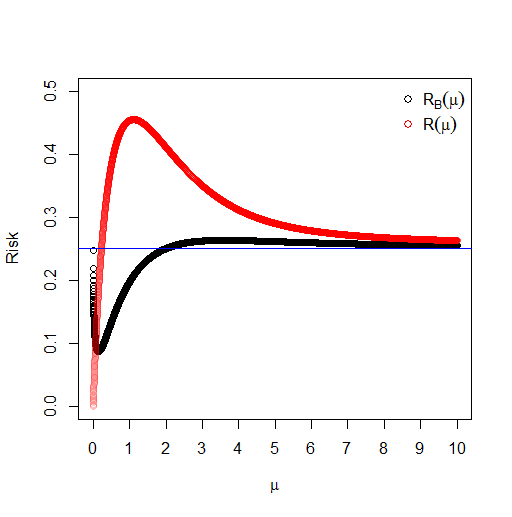}
  \caption{Plots of the Hellinger risk functions $R(\mu)$ and $R_B(\mu)$ for various values of the Poisson parameter $\mu$. The horizontal blue line marks a value of 1/4. See main text for details.}
\label{fig:poissonrate}
\end{figure}

In the last equality we used the fact that under the scaling $n\lambda \rightarrow \mu$, the Binomial random variable converges to a Poisson random variable $K  \sim \text{Po}(\mu)$.  Therefore the risk in this case is given by the function $R(\mu)= \mathbb{E}_{\text{Po}(\mu)} \left[ \left( K^{1/2} -\mu^{1/2}\right)^2 \right] $, where the expectation is taken with respect to the Poisson distribution with parameter $\mu$. If this function was bounded such that $R(\mu) \leq 1/4$, then it would suggest that the standard estimator 
$\hat{\lambda}$ might be globally asymptotically minimax. 
However, plotting the function $R(\mu)$ numerically, we see from Figure \ref{fig:poissonrate} that it attains a maximum value of $\max{R(\mu)} \approx 0.455$ around $\mu=1.11$, and converges to $1/4$ for large values of $\mu$ which corresponds to $\lambda$ away from zero. This shows that the standard estimator doesn't achieve a minimax constant of $1/4$ for all $\lambda$, and illustrates that the difficulty in deriving a minimax rate for the Hellinger distance lies in the Poisson range.
 
As an alternative, we consider the Bayes estimator $\hat{\mu}^{1/2}_B$ for $\mu^{1/2}$, and plot numerically the function $R_B(\mu) :=  \mathbb{E}_{\mu} \left[ ( \hat{\mu}^{1/2}_B - \mu^{1/2} )^2 \right]$. It is known that a conjugate family for the Poisson model is the Gamma family of priors, i.e 
\begin{equation}
\mu \sim f_{\alpha, \beta}(t)= \frac{t^{\alpha-1} \exp{(-t/\beta)}}{\beta^{\alpha} \Gamma(\alpha)} , \ \ \ \ \ t>0
\end{equation}

where $\alpha, \beta$ are the shape and scale parameters respectively. Given an outcome $K=k$ from the Poisson distribution $\text{Po}(\mu)$, the posterior distribution for $\mu$ is easily calculated to be $\Gamma(\alpha+x, \frac{\beta}{\beta+1})$. The Bayes estimator is then the expectation of $\sqrt{\mu}$ taken with respect to the posterior distribution. This is easily calculated to be 
\begin{equation}
\hat{\mu}^{1/2}_B := \frac{\Gamma(x+ \alpha + 1/2)}{\Gamma(x+\alpha)} \left( \frac{\beta}{\beta+1} \right)^{1/2} 
\end{equation}

In Figure \ref{fig:poissonrate}, we plot the risk $R_B(\mu)$ for the Bayes estimate under a particular prior with $\alpha = 0.41$ and $\beta = 200$, and a range of values for $\mu$. We see that the risk remains upper-bounded by a value only slightly greater than 1/4, and for large values of $\mu$ the risk tends to a limiting value of $1/4$. 
This supports the conjecture that the minimax constant for the Hellinger distance is $1/4$.

\section{Quantum Relative Entropy}\label{sec:QRE}

In our derivation of the minimax upper bounds, we bounded the Hellinger risk of estimating the eigenvalues by the Kullback-Leibler (KL) risk for which the asymptotic minimax rate is known. As the KL distance is the classical analogue of the quantum relative entropy $S(\rho \Vert \rho^{\prime}) = {\rm Tr}[ \rho( \log{\rho} - \log{\rho^{\prime}})]$, a question naturally arises - can the techniques used in this paper be applied to derive the minimax rate for the quantum relative entropy (QRE)?  A key element would be to decompose the QRE locally. Similar to decomposition of the Bures distance in (\ref{eq:buresexpansion}), the QRE risk can be shown to locally decompose into a sum of contributions from the KL risk and a term involving the `rotation parameters' . For qubit states, the QRE between the state $\rho$ and the estimate $\hat{\rho}_n$ is represented in terms of the Bloch vectors as \cite{Cortese}
\begin{align}
S(\rho \Vert \hat{\rho}_n) = \frac{1}{2} \bigg[ 
\log{(1-\vert\boldsymbol{r}\vert^2)} - 
\log{(1-\vert\hat{\boldsymbol{r}}_n\vert^2)} 
+ \vert \boldsymbol{r} \vert &\log{ \left( \frac{1+\vert \boldsymbol{r}\vert}{1-\vert\boldsymbol{r}\vert} \right)} \\ \notag
&- \vert \boldsymbol{r} \vert \cos{\Phi} \log{ \left( \frac{1+\vert \hat{\boldsymbol{r}}_n  \vert}{1-\vert \hat{\boldsymbol{r}}_n \vert} \right)} \bigg] 
\end{align}
where $\Phi$ is the angle between the Bloch vectors of the two states. This can be rewritten as 
\begin{eqnarray}
S(\rho \Vert \hat{\rho}_n) &=& D_{KL}(\boldsymbol{\lambda}, \hat{\boldsymbol{\lambda}}_n) +  \frac{1-2\lambda }{2} (1-\cos \Phi) \log{\left( \frac{1-\hat{\lambda}_n}{\hat{\lambda}_n} \right)} 
\nonumber \\
 &=& D_{KL}(\boldsymbol{\lambda}, \hat{\boldsymbol{\lambda}}_n) +  
 \frac{1-2\lambda }{4} (\Phi^2 + O(\Phi^4) ) \log{\left( \frac{1-\hat{\lambda}_n}{\hat{\lambda}_n} 
\right)} 
\label{eq.expansion.re}
\end{eqnarray}
where $D_{KL}(\boldsymbol{\lambda}, \hat{\boldsymbol{\lambda}})$ is the Kullback-Leibler distance between the two distributions $\boldsymbol{\lambda} = (\lambda, 1-\lambda)$ and $\hat{\boldsymbol{\lambda}} =(\hat{\lambda},1-\hat{\lambda})$, and in the second equality we expanded the cosine term to leading order in $\Phi$. The proof of this decomposition can be found in appendix \ref{qreexpand}.

We will show that the global estimator discussed in section \ref{sec:bounds} achieves the rate  $O(n^{-1}\log n)$ and no estimator can achieve faster rates, in particular the `standard' rate $n^{-1}$.

Consider the estimator define by equations \eqref{eq:addbetaKL} and \eqref{eq:case2}, and note that the classical component estimator $\hat{\lambda}_n$ is the minimax optimal estimator for the binomial model and is always larger than $c/n$ for some fixed constant $c>0$. Using the same arguments as in Theorem \ref{th:upperbound} we find 
$$
\sup_\rho \mathbb{E} [D_{KL}(\boldsymbol{\lambda} ,\hat{\boldsymbol{\lambda}}_n) ] = O(n^{-1}).
$$
On the other hand, since $\hat{\lambda}_n \geq c/n$, the second term in equation \eqref{eq.expansion.re} is bounded by $c^\prime \log n  (\Phi^2 + O(\Phi^4))$; since $\Phi$ is estimated at standard rate, the second term is therefore upper bounded as  
$$
\sup_\rho 
\mathbb{E} 
\left[  
\frac{1-2\lambda }{4} (\Phi^2 + O(\Phi^4) ) \log{\left( \frac{1-\hat{\lambda}_n}{\hat{\lambda}_n} 
\right) }\right] 
= 
O(n^{-1} \log n )
$$
which determines the rate.

We will now show that no estimator can have maximum risk converging faster 
that $n^{-1}\log n$. 
Since we are interested in the maximum risk, we will set $\lambda =0$ (pure states), and show that the risk cannot decrease faster that $n^{-1}\log n$ even if we know that the state is pure! As a consequence of local asymptotic normality, any estimator will have the property that 
$\mathbb{P} [ 1- \cos \Phi \geq c/n ] \geq \epsilon $ for some constants $c,\epsilon$. Therefore we will consider the contribution to the risk conditional on $1- \cos \Phi \geq c/n$. By expanding the $D_{KL}(\boldsymbol{\lambda}, \hat{\boldsymbol{\lambda}})$ term we have 
\begin{equation}
S(\rho \Vert \hat{\rho}_n) \geq  
\log \frac{1}{1- \hat{\lambda}_n} + \frac{c}{4n}  \log \left( \frac{1-\hat{\lambda}_n}{\hat{\lambda}_n}\right) . 
\end{equation}
However, the righthand side achieves its minimum at $\hat{\lambda}_n= c/4n$, so the risk is larger 
than $c^\prime \log n /n$. This shows that the minimax risk for the quantum relative entropy scales as $\log n /n$.


%


\section{Conclusion}
 
In this paper we proposed two adaptive estimators for the qubit mixed state, one based on local measurements and the other on collective global measurements. In section \ref{sec:localmeasurement} we upper-bounded the minimax Bures distance risk for the estimator based on local measurements and showed that it scales as $1/n$. In section \ref{sec:globalmeasurement}, we proposed an estimator based on collective measurements and used LAN theory to obtain upper and lower bounds for the risk of $3/2n$ and $5/4n$ respectively. A key element in obtaining the upper bounds was the local decomposition of the Bures risk into contributions from the Hellinger risk of estimating the eigenvalue, and a quadratic contribution from the risk of estimating the `rotation parameters'. While the contribution to the Bures risk from the `rotation terms' is easily shown to scale as $O(1/n)$, we noticed that the difficulty in establishing minimax results for the Bures distance is encapsulated in the challenges of establishing minimax results for the Hellinger risk.  Finally in section \ref{sec:minimax}, we considered these challenges and proposed that a minimax optimal estimator for the mixed qubit state $\rho$ is immediately obtained given a minimax optimal estimator for the binomial parameter under the Hellinger loss function. 

We also briefly considered the derivation of minimax bounds for the quantum relative entropy (QRE) risk. We derived a local decomposition of the QRE similar to the one obtained for the Bures distance, and demonstrated that the global estimator proposed achieves a rate of $O(n^{-1} \log{n})$. We also showed that no estimator can achieve faster rates and established that the minimax QRE risk scales as $O(n^{-1} \log{n})$.

\section*{Acknowledgements} 

The authors would like to thank Michael Nussbaum for insightful and fruitful discussions.

\bibliography{biblio}

\appendix

\section{Appendix}
\subsection{Proof of Lemma \ref{le:localisation}}\label{appendix.1}

The proof is a straightforward application of Hoeffding's inequality.
\begin{theorem}[Hoeffding's inequality]
Let $R_1, \ldots, R_m$ be independent random variables with $a_i \leq R_i \leq b_i$. Let $S= \sum_{i=1}^m R_i$, and $\mu = \mathbb{E}[S]$. Then for all $t>0$
\begin{equation}
\mathbb{P}\left( \vert S - \mu \vert \geq nt \right) \leq 2\exp^{-2n^2t^2/\sum_i(b_i-a_i)}.
\end{equation}
\end{theorem} 

%
We have $X_i \in [-1,1]$, and $\mathbb{E} \sum_i X_i= \frac{\tilde{n}}{3} r_x$. Applying Hoeffding's inequality we get 
\begin{equation}\label{eq:hoeffding}
\mathbb{P}\left( \bigg\vert \frac{3}{\tilde{n}}\sum_{i} X_i -  r_x \bigg\vert^2 \geq t^2 \right) \leq 2\exp^{-2t^2\tilde{n}/3} 
\end{equation}
and similarly for the other spin components. Applying the three inequalities together, with $t^2=n^{2\epsilon_2}-1$ and $\epsilon_2>0$, we have 
\begin{equation}\label{eq:componentbound}
\mathbb{P}\left( \sum_{j=x,y,z} \vert r^{\prime}_j -r_j \vert^2 \geq 3n^{2\epsilon_2-1} \right) \leq 6 \exp^{-2\tilde{n}n^{2\epsilon_2-1}/3} . 
\end{equation}

The estimate $\tilde{\rho}$ is then the closest state in trace distance to the matrix $\frac{1}{2}(\mathbb{1} + \boldsymbol{r}^{\prime} \cdot \boldsymbol{\sigma})$. As $\Vert \rho - \tilde{\rho} \Vert_1^2 = \sum_{i=x,y,z} \vert \tilde{r}_i -r_i \vert^2$, (\ref{eq:componentbound}) implies the stated bound. 
\newline
\qed

\subsection{Expansion of the Bures distance}\label{appendix.2}

Here we derive the expansion (\ref{eq:buresexpansion}) of the Bures distance $D_{B}(\rho, \rho^{\prime})^2 := 2 \left[1-\sqrt{ F(\rho, \rho^{\prime} )} \right]$. We know that for qubits the fidelity between two states can be expressed in terms of the Bloch vectors as
\begin{align*}
F(\rho, \rho^{\prime} ) &:= \frac{1}{2}\left( 1 + \sqrt{1- \vert {\bf r} \vert^2}\sqrt{1-\vert {\bf r}^{\prime} \vert^2} + \mathbf{r} \cdot \mathbf{r}^{\prime} \right) \\
&= \frac{1}{2} \left( 1 +  \sqrt{1- \vert {\bf r} \vert^2}\sqrt{1-\vert  {\bf r}^{\prime} \vert^2} + \vert {\bf r} \vert \vert {\bf r}^{\prime} \vert \cos{\Phi} \right) 
\end{align*}
where $\Phi$ is the angle between the Bloch vectors. For two sufficiently close states, the angle $\Phi$ is small and the cosine term can be expanded as
\begin{align*}
F(\rho, \rho^{\prime} ) &= \frac{1}{2} \left( 1 +  \sqrt{1- \vert {\bf r} \vert^2}\sqrt{1-\vert {\bf r}^{\prime} \vert^2} + \vert {\bf r} \vert \vert {\bf r}^{\prime} \vert - \frac{\vert {\bf r} \vert  \vert {\bf r}^{\prime} \vert}{2} \Phi^2 + \frac{\vert {\bf r} \vert  \vert {\bf r}^{\prime} \vert}{24} \Phi^4 \right) \\
&= \left( \sqrt{(1-\lambda) (1-\lambda^{\prime}) } + \sqrt{\lambda \lambda^{\prime}} \right)^2 - \frac{\vert {\bf r} \vert  \vert {\bf r}^{\prime} \vert}{4} \Phi^2 +  \frac{\vert {\bf r} \vert  \vert {\bf r}^{\prime} \vert}{24} \Phi^4
\end{align*}
where we have use the fact that $\vert {\bf r} \vert = 1-2\lambda$. Therefore the Bures distance  is given by 
\begin{align*}
D_{B}(\rho, \rho^{\prime})^2 &= 2 \left[ 1 - \sqrt{ \left( \sqrt{(1-\lambda) (1-\lambda^{\prime}) } + \sqrt{\lambda \lambda^{\prime}} \right)^2 - \frac{\vert {\bf r }\vert  \vert {\bf r}^{\prime} \vert}{4} \Phi^2 +\frac{\vert {\bf r} \vert  \vert {\bf r}^{\prime} \vert}{24} \Phi^4} \right] \\
&= 2 \left[ 1- \left( \sqrt{(1-\lambda) (1-\lambda^{\prime}) } + \sqrt{\lambda \lambda^{\prime}} \right) + \frac{1}{8} \frac{\vert {\bf r }\vert \vert {\bf r}^{\prime} \vert}{ \sqrt{(1-\lambda) (1-\lambda^{\prime}) } + \sqrt{\lambda \lambda^{\prime}} } \Phi^2 +O(\Phi^4) \right] \\
&= \  D_H(\boldsymbol{\lambda}, \boldsymbol{\lambda}^{\prime})^2 + \frac{1}{4} \frac{ (1-2\lambda)(1-2\lambda^\prime)}{ \sqrt{(1-\lambda) (1-\lambda^{\prime}) } + \sqrt{\lambda \lambda^{\prime}}} \Phi^2 +O(\Phi^4)
\end{align*}  
where $D_H(\boldsymbol{\lambda}, \boldsymbol{\lambda}^{\prime})^2$ is the Hellinger distance between the binary distributions 
$\boldsymbol{\lambda}= (\lambda, 1-\lambda)$ and $\boldsymbol{\lambda}^\prime= (\lambda^\prime, 1-\lambda^\prime)$.
\newline
\qed

\subsection{Proof of Theorem \ref{th:upperbound}}\label{proofofthm} 
Since the first measurement stage is the localisation of the true state by the estimate $\tilde{\rho}$, we write the risk as a sum of two terms 
\begin{align}
R(\rho, \hat{\rho}_n) &= \mathbb{E} \left[ D_B(\rho,\hat{\rho}_n)^2\right] \notag \\
&= \mathbb{P}\left(\vert \tilde{r} \vert \leq \delta \right)  \cdot \mathbb{E}\left[ D_B(\rho,\hat{\rho}_n)^2 ~| ~\vert \tilde{r} \vert \leq \delta \right] + \mathbb{P}\left(\vert \tilde{r} \vert > \delta \right) \cdot \mathbb{E}\left[ D_B(\rho,\hat{\rho}_n)^2 ~ |~ \vert \tilde{r} \vert > \delta\right] 
\end{align}

The expectation is taken over the measurement outcomes given the true state $\rho$. The final estimate $\hat{\rho}_n$ is defined by either (\ref{eq:case1}) or (\ref{eq:case2}) depending on the preliminary estimate $\tilde{\rho}$. Specifically,  if the estimate $\tilde{\rho}$ is within a ball of radius $\delta>0$ of the fully mixed state, we perform standard tomographic measurements on the remaining copies of the state. The final estimate is then the maximum likelihood (ML) estimate given by (\ref{eq:case1}). While in the other instance the technology of LAN is utilised and the final estimate is (\ref{eq:case2}). In order to make the difference between the two estimators explicit, we let $\hat{\rho}^1_n$ denote the final estimator in the case when $\vert \tilde{r} \vert \leq \delta$ and $\hat{\rho}^2_n$ be the final estimate when $\vert \tilde{r} \vert > \delta$. Therefore, we have 
\begin{align}\label{eq:risksum}
R(\rho, \hat{\rho}_n) &=\mathbb{P}\left(\vert \tilde{r} \vert \leq \delta \right)  \cdot \mathbb{E}\left[ D_B(\rho,\hat{\rho}^1_n)^2 \big\vert \ \vert \tilde{r} \vert \leq \delta \right] + \mathbb{P}\left(\vert \tilde{r} \vert > \delta \right) \cdot \mathbb{E}\left[ D_B(\rho,\hat{\rho}^2_n)^2 \big\vert \ \vert \tilde{r} \vert > \delta \right] \notag \\
&=R_1 + R_2.
\end{align}

We consider the contribution to the risk from the term $R_1$ first. Let $B_1$ be a ball of radius $\delta + n^{-1/2+\epsilon_2}$ around the centre of the Bloch sphere. When the true state $\rho \not\in B_1$, we note that the probability $\mathbb{P}\left(\vert \tilde{r} \vert \leq \delta\right)$ goes to zero exponentially fast in $n$. This is because the estimate $\tilde{\rho}$ lies within a ball of radius $O(n^{-1/2 + \epsilon_2})$ around the true state with high probability (Lemma \ref{le:localisation}). This along with the fact that the Bures distance is bounded as $D_B(\sigma, \pi) \leq 2$ for any pair of density matrices $\sigma, \pi$ implies that when $\rho \not\in B_1$, the term $R_1$ can be neglected. However, when $\rho \in B_1$, the term $R_1$ has a non zero contribution and is written as  
\begin{equation}\label{eq:R_1}
R_1 = 
\left\{
\begin{array}{cc}
\mathbb{E}\left[ D_B(\rho,\hat{\rho}^1_n)^2 \big\vert \ \vert \tilde{r} \vert \leq \delta \right] \cdot \mathbb{P}\left(\vert \tilde{r} \vert \leq \delta \right) ,  & \rho \in B_1\\
 o(1) &  \rho \notin B_1
 \end{array}
 \right.
\end{equation}

The term $R_2$ can be  treated similarly. Let $B_2$ be a ball of radius $\delta - n^{-1/2+\epsilon_2}$ around the centre of the Bloch sphere. As the probability $\mathbb{P}(\vert \tilde{r} \vert > \delta )$ decays exponentially if $\rho \in B_2$, the term $R_2$ is relevant only when $\rho \not\in B_2$
\begin{equation}\label{eq:R_2}
\left\{
\begin{array}{cc}
R_2 = \mathbb{E}\left[ D_B(\rho,\hat{\rho}^2_{n})^2 \big\vert \ \vert \tilde{r} \vert > \delta \right] \cdot \mathbb{P}\left(\vert \tilde{r} \vert > \delta \right), & \rho \not\in B_2\\
o(1) & \rho \in B_2
\end{array}
\right.
\end{equation}

Substituting (\ref{eq:R_1}), (\ref{eq:R_2}) in (\ref{eq:risksum}) we see that the minimax risk is bounded from above as  
\begin{equation}
\begin{split}
\limsup \limits_{n\rightarrow \infty} \  \sup_{\rho}\  nR(\rho,\hat{\rho}_n) \leq \limsup \limits_{n\rightarrow \infty} \max \bigg\{ \sup_{\rho \in B_1} & n\mathbb{E}\left[ D_B(\rho,\hat{\rho}^1_n)^2 \big\vert \ \vert \tilde{r} \vert \leq \delta  \right], \\
&\sup_{\rho \not\in B_2} n\mathbb{E}\left[ D_B(\rho,\hat{\rho}^2_n)^2 \big\vert \  \vert \tilde{r} \vert > \delta \right] \bigg\} 
\end{split}
\end{equation}

{\bf Case 1} : $\rho \in B_1$ and $\vert \tilde{r} \vert \leq \delta$

We now evaluate the risk when the state $\rho$ is in $B_1$ while the estimate $\tilde{\rho}$ is within a ball of radius $\delta>0$ around the fully mixed state. 
The final estimate $\hat{\rho}^1_n$ is the ML estimate and given by (\ref{eq:case1}). The outcomes from the $n/3$ repeated measurements in a setting $\sigma_i$ are i.i.d, this implies that the ML estimate of the Bloch vector parameters $r_x, r_y, r_z$ from the outcomes of the standard tomographic measurements are asymptotically Gaussian in distribution 
\begin{equation}
\lim_{n\rightarrow \infty} \sqrt{n}(\hat{r_i}-r_i) = \mathcal{N}(0, I(\rho)^{-1}), \ \ \ \ \ \ \  \ \ \ \ \ \  i=x,y,z
\end{equation}

where the covariance matrix is the inverse of the Fisher information matrix $I(\rho)$. The elements of the matrix $I(\rho)$ are defined for each $i,j \in \{x,y,z \}$ as 
\begin{equation}
I(\rho)_{i,j} 
=\ \frac{1}{( 1+ r_i) (1-r_i )}\  \delta_{i,j} 
\end{equation}

The local expansion of the Bures distance for states away from the boundary of the Bloch sphere is quadratic in the Bloch vector components 
\begin{equation}
D_B(\rho, \hat{\rho}^1_n)^2 =  (\boldsymbol{r} - \hat{\boldsymbol{r}})^{T} G (\boldsymbol{r} - \hat{\boldsymbol{r}}) + O(\Vert \boldsymbol{r} - \hat{\boldsymbol{r}} \Vert^3) 
\end{equation}

where $G$ is the weight matrix of the Bures distance
$$G_{j,k} := \frac{{1}}{4}\left(1+ \frac{r_i^2}{(1-\vert r \vert^2)} \right) \delta_{j,k}.
$$ 
The asymptotic behaviour of the ML estimator (\ref{eq:case1}) together with this local expansion of the Bures distance, implies that the risk of the ML estimate scales as follows for large $n$
\begin{equation}
\mathbb{E} \left[ D_B(\rho, \hat{\rho}^1_n)^2 \right] = \frac{1}{n}{\rm Tr} \left(I(\rho)^{-1}G\right) + o(n^{-1})
\end{equation}

It is easy to see that $I(\rho) \geq \mathbb{1}$, and therefore we have that asymptotically the minimax Bures risk is upper bounded by
\begin{equation}\label{eq:case1risk} 
\limsup \limits_{n\rightarrow \infty} \  \sup_{\rho \in B_1} n\mathbb{E} \left[ D_B(\rho,\hat{\rho}^1_n)^2 \right] 
\leq \frac{3}{4} \left( 1+ \frac{\delta}{1-\delta} \right) 
\end{equation}
where we used the fact that  $\vert r \vert \leq \delta+ O(n^{-1/2 + \epsilon_2})$. 

\vspace{\baselineskip}

\textbf{\textit{Case 2}: $\vert \tilde{r} \vert > \delta$, \ $\rho \not\in B_2$}

We now consider the case when $\tilde{\rho}$ is away from the fully mixed state. Since the state $\tilde{\rho}$ is within the ball of radius $O(n^{-1/2+\epsilon_2})$ of the true state $\rho \not\in B_2$, we consider the local parametrisation of the states $\boldsymbol{\theta}=(\lambda,\boldsymbol{w})$, and perform the secondary measurements on the joint state $\rho^{\boldsymbol{\theta}}_{n} = \rho^{\otimes n}_{\boldsymbol{\theta}}$of $n$ qubits. Using the approximation of the Bures distance in (\ref{eq:buresexpansion}), the risk is expressed as
\begin{align*}
\mathbb{E} \left[ D_B(\rho,\hat{\rho}^2_n)^2  \right] &= \mathbb{E} \left[ D_H(\boldsymbol{\lambda}, \hat{\boldsymbol{\lambda}}_n)^2  + \frac{1}{4} \frac{(1-2\lambda)(1-2\hat{\lambda}_n) }{\left( \sqrt{(1-\lambda)(1-\hat{\lambda}_n)} + \sqrt{\lambda \hat{\lambda}_n} \right)}  \Phi^2 \right] +O(n^{-2})\\
&\leq \mathbb{E} \left[  D_H(\boldsymbol{\lambda}, \hat{\boldsymbol{\lambda}}_n)^2 \right] +\frac{1}{n}\mathbb{E} \left[  (u-\hat{u}_n)^2 + (v - \hat{v}_n)^2 \right] +O(n^{-2}).
\end{align*}

The second term on the right corresponding to the rotation parameters has been evaluated in \cite{GutaJanssens}. Since LAN holds in the limit of large $n$, the problem of estimating the rotation parameters is translated to one of determining the displacement of a Gaussian state $\phi^{\boldsymbol{w}}$. The heterodyne measurement is known to be the optimal measurement in this case \cite{GutaJanssens,GutaKahn}. The estimation of these parameters is described in detail in \cite{GutaJanssens}, and here we only note that both $\mathbb{E} \left[ (u - \hat{u}_n)^2 \right]$ and  $\mathbb{E} \left[ (v - \hat{v}_n)^2 \right]$ are bounded from above by $(1-\lambda)/(2(1-2\lambda)^2) \leq 1/2$. Substituting these values, the minimax risk becomes 
\begin{align}\label{eq:finalrisk}
\limsup \limits_{n\rightarrow \infty} \ \sup_{\rho \not\in B_2} n R(\rho, \hat{\rho}^2_n) &=  \limsup \limits_{n\rightarrow \infty} \ \sup_{\boldsymbol{\theta}} n \mathbb{E} \left[ D_B(\rho,\hat{\rho}^2_n)  \right] \notag \\
&\leq \limsup \limits_{n\rightarrow \infty} \sup_{\lambda} n \mathbb{E} \left[ D_H(\boldsymbol{\lambda}, \hat{\boldsymbol{\lambda}}_n)^2 \right] + 1.
\end{align}
The minimax risk is upper bounded by $1$ plus the minimax risk of the Hellinger distance. We now deal with this term. The expectation is taken over the probability distribution $p_{n,\lambda}(j)$ defined in equation (\ref{eq:blockprob}),  
\begin{eqnarray}
\mathbb{E} \left[ D_H(\boldsymbol{\lambda}, \hat{\boldsymbol{\lambda}}_n)^2 \right] &= &\sum_{j=0,1/2}^{n/2}  D_H(\boldsymbol{\lambda}, \hat{\boldsymbol{\lambda}}_n(j))^2 B_{n,\lambda}(n/2-j) \times K(j,n,\lambda) \nonumber \\
&\leq& \sum_{j=0,1/2}^{n/2}  D_H(\boldsymbol{\lambda}, \hat{\boldsymbol{\lambda}}_n(j))^2 B_{n,\lambda}(n/2-j) \times \vert 1- K(j,n,\lambda) \vert \notag\\
&&+ \sum_{j=0,1/2}^{n/2} D_H(\boldsymbol{\lambda}, \hat{\boldsymbol{\lambda}}_n(j))^2 B_{n,\lambda}(n/2-j) = E_1 + E_2 .\label{eq:hellingerrisk}
 \end{eqnarray}  
We now consider the term $E_1$ separately. We split the sum in $E_1$ over the values of $j \in \mathcal{J}_n$, and $j \not\in \mathcal{J}_n$, where $\mathcal{J}_n$ is interval defined in (\ref{eq:jinterval}). Thus, we have 

\begin{eqnarray}
E_1 &= &\sum_{j\in \mathcal{J}_n} D_H(\boldsymbol{\lambda}, \hat{\boldsymbol{\lambda}}_n(j))^2 B_{n,\lambda}(n/2-j) \times \vert 1- K(j,n,\lambda) \vert \notag \\
&& +\sum_{j\not\in \mathcal{J}_n}  D_H(\boldsymbol{\lambda}, \hat{\boldsymbol{\lambda}}_n(j))^2 B_{n,\lambda}(n/2-j) \times \vert 1- K(j,n,\lambda) \vert \nonumber \\
&\leq& \max_{j \in \mathcal{J}_n}\vert 1- K(j,n,\lambda) \vert \sum_{j\in \mathcal{J}_n} D_H(\boldsymbol{\lambda}, \hat{\boldsymbol{\lambda}}_n(j))^2 B_{n,\lambda}(n/2-j) \nonumber \\
&& + \max_{j \not\in \mathcal{J}_n}\vert 1- K(j,n,\lambda) \vert  \sum_{j\not\in \mathcal{J}_n}  D_H(\boldsymbol{\lambda}, \hat{\boldsymbol{\lambda}}_n(j))^2 B_{n,\lambda}(n/2-j) \nonumber \\
&\leq& O(n^{-1/2+ \epsilon_3}) \sum_{j \in \mathcal{J}_n} D_H(\boldsymbol{\lambda},\hat{\boldsymbol{\lambda}}_n(j))^2 B_{n,\lambda}(n/2-j) 
\nonumber \\
&& + \max_{j \not\in \mathcal{J}_n}\vert 1- K(j,n,\lambda) \vert \sum_{j\not\in \mathcal{J}_n}  2 B_{n,\lambda}(n/2-j) .
\label{eq:e1}
\end{eqnarray} 

In the last inequality, we used the fact that $K(j,n,\lambda) = 1+ O(n^{-1/2 + \epsilon_3})$ on the values of $j \in \mathcal{J}_n$, and that $D_H( \boldsymbol{p}, \boldsymbol{q})^2 \leq 2$ for any pair of probability distributions $\boldsymbol{p}, \boldsymbol{q}$. The value $ \max_{j \not\in \mathcal{J}_n}\vert 1- K(j,n,\lambda) \vert$ is uniformly bounded for $\lambda$ away from 1/2. This along with the fact that the mass of the binomial distribution $B_{n,\lambda}(n/2-j)$ is concentrated on values of $j \in \mathcal{J}_n$ implies that the second term in (\ref{eq:e1}) goes to zero exponentially fast in $n$. Therefore, 
\begin{equation}
E_1 \leq O(n^{-1/2+ \epsilon_3}) \sum_{j \in \mathcal{J}_n} D_H(\boldsymbol{\lambda},\hat{\boldsymbol{\lambda}}_n(k))^2 B_{n,\lambda}(k).
\end{equation}

Substituting this back in (\ref{eq:hellingerrisk}), we have that the Hellinger risk is upper bounded by
\begin{eqnarray}
\mathbb{E} \left[ D_H(\boldsymbol{\lambda},\hat{\boldsymbol{\lambda}}_n)^2 \right] &\leq& O(n^{-1/2+ \epsilon_3}) \sum_{j \in \mathcal{J}_n} D_H(\boldsymbol{\lambda},\hat{\boldsymbol{\lambda}}_n(j))^2 B_{n,\lambda}(n/2-j) \nonumber \\
&&+ \sum_{j=0,1/2}^{n/2} D_H(\boldsymbol{\lambda}, \hat{\boldsymbol{\lambda}}_n(j))^2 B_{n,\lambda}(n/2-j) \nonumber\\
&\leq &O(n^{-1/2+ \epsilon_3}) \sum_{k=0}^n D_H(\boldsymbol{\lambda},\hat{\boldsymbol{\lambda}}_n(k))^2 B_{n,\lambda}(k) + \sum_{k=0}^{n} D_H(\boldsymbol{\lambda}, \hat{\boldsymbol{\lambda}}_n(k))^2 B_{n,\lambda}(k) \nonumber \\
& =&  \left(1+O(n^{-1/2+ \epsilon_3})\right) \mathbb{E}_{\text{Binom}} \left[ D_{H}(\boldsymbol{\lambda},\hat{\boldsymbol{\lambda}}_n(k))^2 \right] \label{eq:hellingerbinom} \\
& \leq& \left(1+O(n^{-1/2+ \epsilon_3})\right) \mathbb{E}_{\text{Binom}} \left[ D_{KL}(\boldsymbol{\lambda},\hat{\boldsymbol{\lambda}}_n(k)) \right]. \label{eq:KLbinom}
\end{eqnarray}
In the second line we expand the sums over the entire support of the binomial distribution $k \in \{0, \ldots, n \}$, and let $\mathbb{E}_{\text{Binom}}$ mark expectation with respect to the binomial distribution. The last inequality employs the inequality between the squared Hellinger distance and the Kullback-Leibler (KL) distance, defined as $D_{KL}(\boldsymbol{p}, {\boldsymbol{q}}) := p \log{\frac{p}{q}} + (1-p) \log{\frac{1-p}{1-q}}$. Substituting (\ref{eq:KLbinom}) in (\ref{eq:finalrisk}), and noting that the `add-beta' estimator $\hat{\lambda}_n$ was defined in (\ref{eq:addbetaKL}) over the full support, we have
$$
\limsup \limits_{n \rightarrow \infty} \sup_{\rho \not\in B_2} n R(\rho, \hat{\rho}^2_n) \leq \limsup \limits_{n \rightarrow \infty} \sup_{\lambda} n \mathbb{E}_{\text{Binom}} \left[ D_{KL}(\boldsymbol{\lambda},\hat{\boldsymbol{\lambda}}_n) \right]  \left(1+O(n^{-1/2+ \epsilon_3}) \right) + 1 
\leq   \frac{1}{2}+ 1.
$$

The rate follows from the fact that the `add beta' estimator of the binomial parameter, defined in equation (\ref{eq:addbetaKL}), is known to be asymptotically minimax for the KL risk, achieving the rate $\frac{1}{2n}(1+o(1))$ \cite{BraessDette,BraessSauer}.

Comparing this rate with the one in (\ref{eq:case1risk}), we see that $3/2$ is the larger value provided $\delta < 1/2$. This gives the upper bound stated in Theorem \ref{th:upperbound}. 
\newline
\qed   

\subsection{Expansion of Quantum Relative Entropy}\label{qreexpand}  

\noindent We derive the expansion of the quantum relative entropy $S(\rho \Vert \rho^{\prime}) = {\rm Tr}[ \rho( \log{\rho} - \log{\rho^{\prime}})]$. For qubits the relative entropy between two states can be expressed in terms of the Bloch vector components as 
\begin{align}
S(\rho \Vert \rho^{\prime}) = \frac{1}{2} \bigg[ \log{(1-\vert\boldsymbol{r}\vert^2)} - \log{(1-\vert\boldsymbol{r}^{\prime}\vert^2)} + \vert \boldsymbol{r} \vert &\log{ \left( \frac{1+\vert \boldsymbol{r}\vert}{1-\vert\boldsymbol{r}\vert} \right)} \\ \notag
&- \vert \boldsymbol{r} \vert \cos{\Phi} \log{ \left( \frac{1+\vert \boldsymbol{r}^{\prime} \vert}{1-\vert \boldsymbol{r}^{\prime} \vert} \right)} \bigg] 
\end{align}

\noindent where $\Phi$ is the angle between the Bloch vectors of the two states. For sufficiently close states, the angle $\Phi$ is small and the cosine term in the above equation can be expanded as 

\begin{align}
S(\rho \Vert \rho^{\prime}) = \frac{1}{2} \bigg[ \log{(1-\vert\boldsymbol{r}\vert^2)} &- \log{(1-\vert\boldsymbol{r}^{\prime}\vert^2)} + \vert \boldsymbol{r} \vert \log{ \left( \frac{1+\vert \boldsymbol{r} \vert}{1- \vert \boldsymbol{r} \vert} \right)} \\
&- \vert \boldsymbol{r} \vert  \log{ \left( \frac{1+\vert \boldsymbol{r}^{\prime} \vert}{1-\vert \boldsymbol{r}^{\prime} \vert} \right)} \bigg] + \frac{\vert \boldsymbol{r} \vert}{4} (\Phi^2 +O(\Phi^4)) \log{ \left( \frac{1+\vert \boldsymbol{r}^{\prime} \vert}{1-\vert \boldsymbol{r}^{\prime} \vert} \right)} \notag
\end{align}

\noindent Using the fact that $\vert \boldsymbol{r} \vert = 1-2\lambda $ and simplifying, we get
 
\begin{align}
S(\rho \Vert \rho^{\prime}) &= \frac{1}{2} \bigg[ \log{\left(\frac{\lambda(1-\lambda)}{\lambda^{\prime}(1-\lambda^{\prime})} \right)} - (1-2\lambda) \log{ \left( \frac{\lambda(1-\lambda^{\prime})}{\lambda^{\prime}(1-\lambda)} \right)} \bigg]  + \frac{\vert \boldsymbol{r} \vert}{4} (\Phi^2+O(\Phi^4)) \log{ \left( \frac{1+\vert \boldsymbol{r}^{\prime} \vert}{1-\vert \boldsymbol{r}^{\prime} \vert} \right)}  \notag \\ 
&=  \lambda \log{\left( \frac{\lambda}{\lambda^{\prime}} \right)} + (1-\lambda) \log{\left( \frac{1-\lambda}{1-\lambda^{\prime}} \right)} + \frac{1-2\lambda}{4} (\Phi^2+O(\Phi^4)) \log{\left( \frac{1-\lambda^{\prime}}{\lambda^{\prime}} \right)} \\ 
&= D_{KL}(\boldsymbol{\lambda}, \boldsymbol{\lambda^{\prime}}) +  \frac{1-2\lambda }{4} (\Phi^2 +O(\Phi^4)) \log{\left( \frac{1-\lambda^{\prime}}{\lambda^{\prime}} \right)}
\end{align}

\noindent Where $D_{KL}(\boldsymbol{\lambda}, \boldsymbol{\lambda^{\prime}})$ is the Kullback-Leibler distance between the two distributions $\boldsymbol{\lambda} = (\lambda, 1-\lambda)$ and $\boldsymbol{\lambda^{\prime}} =(\lambda^{\prime},1-\lambda^{\prime})$.  
\newline
\qed 

\end{document}